\documentclass[12pt]{article}
\usepackage{graphicx}
\usepackage{dcolumn}
\usepackage{amsfonts}
\usepackage[colorlinks,citecolor=blue,urlcolor=blue]{hyperref}
\usepackage{amssymb}
\usepackage{mathrsfs}
\usepackage{graphicx}
\usepackage{amsmath}
\usepackage[usenames]{color}
\usepackage{bm, times, graphics}
\usepackage{rotating}
\usepackage{epstopdf}
\usepackage{epsfig}
\usepackage{url}
\usepackage[default]{jasa_harvard}
\usepackage{JASA_manu}

\newtheorem{theorem}{Theorem}
\newtheorem{corollary}{Corollary}
\newtheorem{proposition}{Proposition}
\newtheorem{lemma}{Lemma}
\newtheorem{example}{Example}
\newtheorem{remark}{Remark}
\newtheorem{definition}{Definition}
\newcommand{\beq}{\begin{equation}}
\newcommand{\eeq}{\end{equation}}
\newcommand{\beas}{\begin{eqnarray*}}
\newcommand{\eeas}{\end{eqnarray*}}
\newcommand{\bea}{\begin{eqnarray}}
\newcommand{\eea}{\end{eqnarray}}
\newcommand{\bei}{\begin{itemize}}
\newcommand{\eei}{\end{itemize}}
\newcommand{\ben}{\begin{enumerate}}
\newcommand{\een}{\end{enumerate}}
\newcommand{\bet}{\begin{theorem}}
\newcommand{\eet}{\end{theorem}}
\newcommand{\bel}{\begin{lemma}}
\newcommand{\eel}{\end{lemma}}
\newcommand{\bep}{\begin{proposition}}
\newcommand{\eep}{\end{proposition}}
\newcommand{\bed}{\begin{definition}}
\newcommand{\eed}{\end{definition}}
\newcommand{\bec}{\begin{corollary}}
\newcommand{\eec}{\end{corollary}}
\newcommand{\bex}{\begin{example}}
\newcommand{\eex}{\end{example}}

\input{tcilatex}

\begin{document}

\title{Asymptotically Normal and Efficient Estimation of Covariate-Adjusted
Gaussian Graphical Model}
\author{Mengjie Chen, \\
Program of Computational Biology and Bioinformatics, \\
\\
Zhao Ren, \\
Department of Statistics, \\
\\
Hongyu Zhao, \\
Department of Biostatistics, School of Public Health,\\
\\
Harrison Zhou\\
Department of Statistics,\\
Yale University, New Haven, CT 06520, USA.\\
Email: harrison.zhou@yale.edu. }
\maketitle


\newpage

\begin{center}
\textbf{Abstract}
\end{center}

A tuning-free procedure is proposed to estimate the covariate-adjusted
Gaussian graphical model. For each finite subgraph, this estimator is
asymptotically normal and efficient. As a consequence, a confidence interval
can be obtained for each edge. The procedure enjoys easy implementation and
efficient computation through parallel estimation on subgraphs or edges. We
further apply the asymptotic normality result to perform support recovery
through edge-wise adaptive thresholding. This support recovery procedure is
called ANTAC, standing for Asymptotically Normal estimation with
Thresholding after Adjusting Covariates. ANTAC outperforms other
methodologies in the literature in a range of simulation studies. We apply
ANTAC to identify gene-gene interactions using an eQTL dataset. Our result
achieves better interpretability and accuracy in comparison with CAMPE.

\vspace*{.3in}


\newpage

\section{Introduction}

\label{sec:intro}

Graphical models have been successfully applied to a broad range of studies
that investigate the relationships among variables in a complex system. With
the advancement of high-throughput technologies, an unprecedented amount of
features can be collected for a given system. Therefore, the inference with
graphical models has become more challenging. To better understand the
complex system, novel methods under high dimensional setting are extremely
needed. Among graphical models, Gaussian graphical models have recently
received considerable attention for their applications in the analysis of
gene expression data. It provides an approach to discover and analyze gene
relationships, which offers insights into gene regulatory mechanism. However
gene expression data alone are not enough to fully capture the complexity of
gene regulation. Genome-wide expression quantitative trait loci (eQTL)
studies, which simultaneously measure genetic variation and gene expression
levels, reveal that genetic variants account for a large proportion of the
variability of gene expression across different individuals \cite%
{rockman2006genetics}. Some genetic variants may confound the genetic
network analysis, thus ignoring the influence of them may lead to false
discoveries. Adjusting the effect of genetic variants is of importance for
the accurate inference of genetic network at the expression level. A few
papers in the literature have considered to accommodate covariates in
graphical models. See, for example, \citeasnoun{li2012sparse}, %
\citeasnoun{yin2013adjusting} and \citeasnoun{cai2013covariate} introduced
Gaussian graphical model with adjusted covariates, and %
\citeasnoun{cheng2012sparse} introduced additional covariates to Ising
models.

This problem has been naturally formulated as joint estimation of the
multiple regression coefficients and the precision matrix in Gaussian
settings. Since it is widely believed that genes operate in biological
pathways, the graph for gene expression data is expected to be sparse. Many
regularization-based approaches have been proposed in the literature. Some
use a joint regularization penalty for both the multiple regression
coefficients and the precision matrix and solve iteratively \cite%
{obozinski2011support,yin2011sparse,peng2010regularized}. Others apply a
two-stage strategy: estimating the regression coefficients in the first
stage and then estimating the precision matrix based on the residuals from
the first stage. For all these methods, the thresholding level for support
recovery depends on the unknown matrix $l_{1}$ norm of the precision matrix
or an irrepresentable condition on the Hessian tensor operator, thus those
theoretically justified procedures can not be implemented practically. In
practice, the thresholding level is often selected through cross-validation.
When the dimension $p$ of the precision matrix is relatively high,
cross-validation is computationally intensive, with a jeopardy that the
selected thresholding level is very different from the optimal one. As we
show in the simulation studies presented in Section \ref{sec:simu}, the
thresholding levels selected by the cross-validation tend to be too small,
leading to an undesired denser graph estimation in practice. In addition,
for current methods in the literature, the thresholding level for support
recovery is set to be the same for all entries of the precision matrix,
which makes the procedure non-adaptive.

In this paper, we propose a tuning free methodology for the joint estimation
of the regression coefficients and the precision matrix. The estimator for
each entry of the precision matrix or each partial correlation is
asymptotically normal and efficient. Thus a P-value can be obtained for each
edge to reflect the statistical significance of each entry. In the gene
expression analysis, the P-value can be interpreted as the significance of
the regulatory relationships among genes. This method is easy to implement
and is attractive in two aspects. First, it has the scalability to handle
large datasets. Estimation on each entry is independent and thus can be
parallelly computed. As long as the capacity of instrumentation is adequate,
those steps can be distributed to accommodate the analysis of high
dimensional data. Second, it has the modulability to estimate any subgraph
with special interests. For example, biologists may be interested in the
interaction of genes play essential roles in certain biological processes.
This method allows them to specifically target the estimation on those
genes. An R package implementing our method has been developed and is
available on the CRAN website.

We apply the asymptotic normality and efficiency result to do support
recovery by edge-wise adaptive thresholding. This rate-optimal support
recovery procedure is called ANTAC, standing for Asymptotically Normal
estimation with Thresholding after Adjusting Covariates. This work is
closely connected to a growing literature on optimal estimation of large
covariance and precision matrices. Many regularization methods have been
proposed and studied. For example, Bickel and Levina \cite{BL08A,BL08B}
proposed banding and thresholding estimators for estimating bandable and
sparse covariance matrices respectively and obtained rate of convergence for
the two estimators. See also \citeasnoun{ELK08} and \citeasnoun{LFAN09}. %
\citeasnoun{CZZH10} established the optimal rates of convergence for
estimating bandable covariance matrices. \citeasnoun{CZh12Sparse} and %
\citeasnoun{CLZ12} obtained the minimax rate of convergence for estimating
sparse covariance and precision matrices under a range of losses including
the spectral norm loss. Most closely related to this paper is the work in %
\citeasnoun{RenSunZhZh2013} where fundamental limits were given for
asymptotically normal and efficient estimation of sparse precision matrices.
Due to the complication of the covariates, the analysis in this paper is
more involved.

We organize the rest of the paper as follows. Section \ref{sec: Model and
Methodology} describes the covariate-adjusted Gaussian graphical model and
introduces our novel two-step procedure. Corresponding theoretical studies
on asymptotic normal distribution and adaptive support recovery are
presented in Sections \ref{sec: Asymp Normal}-\ref{sec: model selection}.
Simulation studies are carried out in Section \ref{sec:simu}. Section \ref%
{se: real data} presents the analysis of eQTL data. Proofs for theoretical
results are collected in Section \ref{se: proof of main theorems}. We
collect a key lemma and auxiliary results for proving the main results in
Section \ref{se: key lemma} and Appendix \ref{se: appdix}.

\section{Covariate-adjusted Gaussian Graphical Model and Methodology}

\label{sec: Model and Methodology}

In this section we first formally introduce the covariate-adjusted Gaussian
graphical model, and then propose a two-step procedure for estimation of the
model.

\subsection{Covariate-adjusted Gaussian Graphical Model}

Let $\left( X^{(i)},Y^{(i)}\right) $, $i=1,...,n$, be i.i.d. with%
\begin{equation}
Y^{(i)}=\Gamma _{p\times q}X^{(i)}+Z^{\left( i\right) }\text{,}
\label{Model}
\end{equation}%
where $\Gamma $ is a $p\times q$ unknown coefficient matrix, and $Z^{\left(
i\right) }$ is a $p-$dimensional random vector following a multivariate
Gaussian distribution $N\left( 0,\Omega ^{-1}\right) $ and is independent of
$X^{(i)}$. For the genome-wide expression quantitative trait studies, $%
Y^{(i)}$ is the observed expression levels for $p$ genes of the $i-$th
subject and $X^{(i)}$ is the corresponding values of $q$ genetic markers. We
will assume that $\Omega $ and $\Gamma _{p\times q}$ are sparse. The
precision matrix $\Omega $ is assumed to be sparse partly due to the belief
that genes operate in biological pathways, and the sparseness structure of $%
\Gamma $ reflects the sensitivity of confounding of genetic variants in the
genetic network analysis.

We are particularly interested in the graph structure of random vector $%
Z^{\left( i\right) }$, which represents the genetic networks after removing
the effect of genetic markers. Let $G=(V,E)$ be an undirected graph
representing the conditional independence relations between the components
of a random vector $Z^{\left( 1\right) }=(Z_{11},\dotsc ,Z_{1p})^{T}$. The
vertex set $V=\{V_{1},\dotsc ,V_{p}\}$ represents the components of $Z$. The
edge set $E$ consists of pairs $(i,j)$ indicating the conditional dependence
between $Z_{1i}$ and $Z_{1j}$ given all other components. In the genetic
network analysis, the following question is fundamental: Is there an edge
between $V_{i}$ and $V_{j}$? It is well known that recovering the structure
of an undirected Gaussian graph $G=\left( V,E\right) $ is equivalent to
recovering the support of the population precision matrix $\Omega =\left(
\omega _{ij}\right) $ of the data in the Gaussian graphical model. There is
an edge between $V_{i}$ and $V_{j}$, i.e., $(i,j)\in E$, if and only if $%
\omega _{ij}\neq 0$. See, for example, \citeasnoun{GraMod}. Consequently,
the support recovery of the precision matrix $\Omega $ yields the recovery
of the structure of the graph $G$.

Motivated by biological applications, we consider the high-dimensional case
in this paper, allowing the dimension to exceed or even be far greater than
the sample size, $\min \left\{ p,q\right\} \geq n$. The main goal of this
work is not only to provide a fully data driven and easily implementable
procedure to estimate the network for the covariate-adjusted Gaussian
graphical model, but also to provide a confidence interval for estimation of
each entry of the precision matrix $\Omega$.

\subsection{A Two-step Procedure}

In this section, we propose a two-step procedure to estimate $\Omega $. In
the first step of the two-step procedure, we apply a scaled lasso method to
obtain an estimator $\hat{\Gamma}=\left( \hat{\gamma}_{1},\ldots ,\hat{\gamma%
}_{p}\right) ^{T}$ of $\Gamma $. This procedure is tuning free. This is
different from other procedures in the literature for the sparse linear
regression, such as standard lasso and Dantzig selector which select tuning
parameters by cross-validation and can be computationally very intensive for
high dimensional data. In the second step, we approximate each $Z^{(i)}$ by $%
\hat{Z}^{(i)}=Y^{(i)}-\hat{\Gamma}_{p\times q}X^{(i)}$, then apply the
tuning-free methodology proposed in \citeasnoun{RenSunZhZh2013} for the
standard Gaussian graphical model to estimate each entry $\omega _{ij}$ of $%
\Omega $, pretending that each $\hat{Z}^{(i)}$ was $Z^{(i)}$. As a
by-product, we have an estimator $\hat{\Gamma}$ of $\Gamma $, which is shown
to be rate optimal under different matrix norms, however our main goal is
not to estimate $\Gamma$, but to make inference on $\Omega$.

\subsubsection{Step 1}

Denote the $n$ by $q$ dimensional explanatory matrix by $\mathbf{X}=\left(
X^{(1)},\ldots ,X^{(n)}\right) ^{T}$,\textbf{\ }where the $i$th row of
matrix is from the $i-$th sample $X^{(i)}$. Similarly denote the $n$ by $p$
dimensional response matrix by $\mathbf{Y}=\left( Y^{(1)},\ldots
,Y^{(n)}\right) ^{T}$ and the noise matrix by $\mathbf{Z}=\left(
Z^{(1)},\ldots ,Z^{(n)}\right) ^{T}$. Let $\mathbf{Y}_{j}$ and $\mathbf{Z}%
_{j}$ be the $j-$th column of $\mathbf{Y}$ and $\mathbf{Z}$ respectively.
For each $j=1,...,p$, we apply a scaled lasso penalization to the univariate
linear regression of $\mathbf{Y}_{j}$ against $\mathbf{X}$ as follows,
\begin{equation}
\text{\textbf{Step} }\mathbf{1}:\left\{ \hat{\gamma}_{j},\hat{\sigma}%
_{jj}^{1/2}\right\} =\arg \min_{b\in \mathbb{R}^{q},\theta \in \mathbb{R}%
^{+}}\left\{ \frac{\left\Vert \mathbf{Y}_{j}-\mathbf{X}b\right\Vert ^{2}}{%
2n\theta }+\frac{\theta }{2}+\lambda _{1}\sum_{k=1}^{q}\frac{\left\Vert
\mathbf{X}_{k}\right\Vert }{\sqrt{n}}\left\vert b_{k}\right\vert \right\}
\text{,}  \label{first step: scaled lasso}
\end{equation}%
where the weighted penalties are chosen to be adaptive to each variance $%
Var\left( X_{1k}\right) $ such that an explicit value can be given for the
parameter $\lambda _{1}$, for example, one of the theoretically justified
choices is $\lambda _{1}=\sqrt{2(1+\frac{\log p}{\log q})/n}$. The scaled
lasso (\ref{first step: scaled lasso}) is jointly convex in $b$ and $\theta $%
. The global optimum can be obtained through alternatively updating between $%
b$ and $\theta $. The computational cost is nearly the same as that of the
standard lasso. For more details about its algorithm, please refer to %
\citeasnoun{SZH12-ScaledLaoo}.

Define the estimate of \textquotedblleft noise\textquotedblright\ $\mathbf{Z}%
_{j}$ as the residue of the scaled lasso regression by%
\begin{equation}
\mathbf{\hat{Z}}_{j}=\mathbf{Y}_{j}-\mathbf{X}\hat{\gamma}_{j}\text{,}
\label{first step: estimator of Z}
\end{equation}%
which will be used in the second step to make inference for $\Omega $.

\subsubsection{Step 2}

In the second step, we propose a tuning-free regression approach to estimate
$\Omega $ based on $\mathbf{\hat{Z}}$ defined in Equation (\ref{first step:
estimator of Z}), which is different from other methods proposed in the
literature, including \citeasnoun{cai2013covariate} or %
\citeasnoun{yin2013adjusting}. An advantage of our approach is the ability
to provide an asymptotically normal and efficient estimation of each entry
of the precision matrix $\Omega $.

We first introduce some convenient notations for a subvector or a submatrix.
For any index subset $A$ of $\left\{ 1,2,\ldots ,p\right\} $ and a vector $W$
of length $p$, we use $W_{A}$ to denote a vector of length $\left\vert
A\right\vert $ with elements indexed by $A$. Similarly for a matrix $U$ and
two index subsets $A$ and $B$ of $\left\{ 1,2,\ldots ,p\right\} $, we can
define a submatrix $U_{A,B}$ of size $|A|\times |B|$ with rows and columns
of $U$ indexed by $A$ and $B$, respectively. Let $W=(W_{1},\dotsc ,W_{p})^{T}
$, representing each $Z^{(i)}$, follow a Gaussian distribution $N\left(
0,\Omega ^{-1}\right) $. It is well known that
\begin{equation}
W_{A}|W_{A^{c}}=N\left( -\Omega _{A,A}^{-1}\Omega _{A,A^{c}}W_{A^{c}},\Omega
_{A,A}^{-1}\right) \text{.}  \label{CondiDis2}
\end{equation}%
For $A=\left\{ i,j\right\} $, equivalently we may write
\begin{equation}
\left( W_{i},W_{j}\right) =W_{\left\{ i,j\right\} ^{c}}^{T}\mathbf{\beta }%
+\left( \eta _{i},\eta _{j}\right) \text{,}  \label{regression}
\end{equation}%
where the coefficients and error distributions are%
\begin{equation}
\mathbf{\beta }=-\Omega _{A^{c},A}\Omega _{A,A}^{-1},\text{ \ }\left( \eta
_{i},\eta _{j}\right) ^{T}\sim N\left( 0,\Omega _{A,A}^{-1}\right) \text{.}
\label{true beta}
\end{equation}%
Based on the regression interpretation (\ref{regression}), we have the
following data version of the multivariate regression model%
\begin{equation}
\mathbf{Z}_{A}=\mathbf{Z}_{A^{c}}\mathbf{\beta }+\mathbf{\epsilon }_{A}\text{%
,}  \label{regression-data}
\end{equation}%
where $\mathbf{\beta }$ is a $\left( p-2\right) $ by $2$ dimensional
coefficient matrix. If we know $\mathbf{Z}_{A}$ and $\mathbf{\beta }$, an
asymptotically normal and efficient estimator of $\Omega _{A,A}$ is $\left(
\mathbf{\epsilon }_{A}^{T}\mathbf{\epsilon }_{A}/n\right) ^{-1}$.

But of course $\mathbf{\beta }$ is unknown and we only have access to the
estimated observations $\mathbf{\hat{Z}}$ from Equation (\ref{first step:
estimator of Z}). We replace $\mathbf{Z}_{A}$ and $\mathbf{Z}_{A^{c}}$ by $%
\mathbf{\hat{Z}}_{A}$ and $\mathbf{\hat{Z}}_{A^{c}}$ respectively in the
regression (\ref{regression-data}) to estimate $\mathbf{\beta }$ as follows.
For each $m\in A=\left\{ i,j\right\} $, we apply a scaled lasso penalization
to the univariate linear regression of $\mathbf{\hat{Z}}_{m}$ against $%
\mathbf{\hat{Z}}_{A^{c}}$,
\begin{equation}
\text{\textbf{Step} }\mathbf{2}:\left\{ \mathbf{\hat{\beta}}_{m},\hat{\psi}%
_{mm}^{1/2}\right\} =\arg \min_{b\in \mathbb{R}^{p-2},\sigma \in \mathbb{R}%
^{+}}\left\{ \frac{\left\Vert \mathbf{\hat{Z}}_{m}-\mathbf{\hat{Z}}%
_{A^{c}}b\right\Vert ^{2}}{2n\sigma }+\frac{\sigma }{2}+\lambda
_{2}\sum_{k\in A^{c}}\frac{\left\Vert \mathbf{\hat{Z}}_{k}\right\Vert }{%
\sqrt{n}}\left\vert b_{k}\right\vert \right\} \text{,}
\label{Second step: scaled Lasso}
\end{equation}%
where the vector $b$ is indexed by $A^{c}$, and one of the theoretically
justified choices of $\lambda _{2}$ is $\lambda _{2}=\sqrt{\frac{2\log p}{n}}
$. Denote the residuals of the scaled lasso regression by%
\begin{equation}
\mathbf{\hat{\epsilon}}_{A}=\mathbf{\hat{Z}}_{A}-\mathbf{\hat{Z}}_{A^{c}}%
\mathbf{\hat{\beta}}\text{,}  \label{epsilonhat}
\end{equation}%
and then define
\begin{equation}
\hat{\Omega}_{A,A}=\left( \mathbf{\hat{\epsilon}}_{A}^{T}\mathbf{\hat{%
\epsilon}}_{A}/n\right) ^{-1}\text{.}  \label{estimator2}
\end{equation}

This extends the methodology proposed in \citeasnoun{RenSunZhZh2013} for
Gaussian graphical model to corrupted observations. The approximation error $%
\mathbf{\hat{Z}-Z}$ affects inference for $\Omega $. Later we show if $%
\Gamma $ is sufficient sparse, $\Gamma _{p\times q}X$ can be well estimated
so that the approximation error is negligible. When both $\Omega $ and $%
\Gamma $ are sufficiently sparse, $\hat{\Omega}_{A,A}$ can be shown to be
asymptotically normal and efficient. An immediate application of the
asymptotic normality result is to perform adaptive graphical model selection
by explicit entry-wise thresholding, which yields a rate-optimal adaptive
estimation of the precision matrix $\Omega $ under various matrix $l_{w}$
norms. See Theorems \ref{theorem:step2}, \ref{support recovery} and \ref%
{matrixnorm} in Section \ref{sec: Asymp Normal} and \ref{sec: model
selection} for more details.

\section{Asymptotic Normality Distribution of the Estimator}

\label{sec: Asymp Normal}

In this section we first give theoretical properties of the estimator $%
\mathbf{\hat{Z}}$ as well as $\hat{\Gamma}$, then present the asymptotic
normality and efficiency result for estimation of $\Omega $.

We assume the coefficient matrix $\Gamma $ is sparse, and entries of $X$
with mean zero are bounded since the gene marker is usually bounded.

\begin{enumerate}
\item[1.] The coefficient matrix $\Gamma $ satisfies the following sparsity
condition,%
\begin{equation}
\max_{i}\Sigma _{j\neq i}\min \left\{ 1,\frac{\left\vert \gamma
_{ij}\right\vert }{\lambda _{1}}\right\} =s_{1}\text{,}
\label{sparsity definition gamma}
\end{equation}%
where in this paper $\lambda _{1}$ is at an order of $\sqrt{\frac{\log q}{n}}
$. Note that $s_{1}\leq \max_{i}\Sigma _{j\neq i}I\left\{ \gamma _{ij}\neq
0\right\} $, the maximum of the exact row sparseness among all rows of $%
\Gamma $.

\item[2.] There exist positive constants $M_{1}$ and $M_{2}$ such that $%
1/M_{1}\leq \lambda _{\min }\left( Cov(X^{(1)})\right) $ and $1/M_{2}\leq
\lambda _{\min }\left( \Omega \right) \leq \lambda _{\max }\left( \Omega
\right) \leq M_{2}$.

\item[3.] There is a constant $B>0$ such that
\begin{equation}
|X_{ij}|\ \leq B\text{ for all }i\text{ and }j\text{.}
\label{subGau:element}
\end{equation}
\end{enumerate}

It is worthwhile to note that the boundedness assumption (\ref%
{subGau:element}) does not imply the $X^{(1)}$ is jointly sub-gaussian,
i.e., $X^{(1)}$ is allowed to be not jointly sub-gaussian as long as above
conditions are satisfied. In the high dimensional regression literature, it
is common to assume the joint sub-gaussian condition on the design matrix as
follows,

\begin{enumerate}
\item[3'.] We shall assume that the distribution of $X^{(1)}$ is jointly
sub-gaussian with parameter $\left( M_{1}\right) ^{1/2}>0$ in the sense that
\begin{equation}
\mathbb{P}\{|v^{T}X^{(1)}|>t\}\leq e^{-t^{2}/2M_{1}}\text{ for all }t>0\text{
and }\Vert v\Vert _{2}=1\text{.}  \label{subGau}
\end{equation}
\end{enumerate}

We analyze the Step 1 of the procedure in Equation (\ref{first step: scaled
lasso}) under Conditions 1-3 as well as Conditions 1-2 and 3'. The optimal
rates of convergence are obtained under the matrix $l_{\infty }$ norm and
Frobenius norm for estimation of $\Gamma $, which yield a rate of
convergence for estimation of each $\mathbf{Z}_{j}$ under the $l_{2}$ norm.

\begin{theorem}
\label{theorem:step1} Let $\lambda _{1}=\left( 1+\varepsilon _{1}\right)
\sqrt{\frac{2\delta _{1}\log q}{n}}$ for any $\delta _{1}\geq 1$ and $%
\varepsilon _{1}>0$ in Equation (\ref{first step: scaled lasso}). Assume
that
\begin{equation*}
s_{1}=o\left( \min \left\{ \frac{n}{\log ^{3}n\log q},\sqrt{\frac{n}{\log q}}%
\right\} \right) \text{.}
\end{equation*}%
Under Conditions 1-3 we have%
\begin{eqnarray}
\mathbb{P}\left\{ \frac{1}{n}\left\Vert \mathbf{\hat{Z}}_{j}-\mathbf{Z}%
_{j}\right\Vert ^{2}>C_{1}s_{1}\frac{\log q}{n}\right\}  &\leq &o\left(
q^{-\delta _{1}+1}\right) \text{ for each }j\text{,}
\label{First step: z bound} \\
\mathbb{P}\left\{ \left\Vert \mathbf{\hat{\Gamma}}-\mathbf{\Gamma }%
\right\Vert _{l_{\infty }}>C_{2}s_{1}\sqrt{\frac{\log q}{n}}\right\}  &\leq
&o\left( p\cdot q^{-\delta _{1}+1}\right) \text{,}
\label{First step: gamma Lsup bound} \\
\mathbb{P}\left\{ \frac{1}{p}\left\Vert \mathbf{\hat{\Gamma}}-\mathbf{\Gamma
}\right\Vert _{F}^{2}>C_{3}s_{1}\frac{\log q}{n}\right\}  &\leq &o\left(
p\cdot q^{-\delta _{1}+1}\right) \text{.}
\label{First step: gamma Fro bound}
\end{eqnarray}%
Moreover, if we replace Condition 3 by the weaker version Condition 3', all
results above still hold under a weaker assumption on $s_{1}$,
\begin{equation}
s_{1}=o\left( \frac{n}{\log q}\right) \text{.}  \label{sparsity condition s1}
\end{equation}
\end{theorem}

The proof of Theorem \ref{theorem:step1} is provided in the Section \ref{se:
proof of theorem 1}.

\begin{remark}
Under the assumption that the $l_{r}$ norm of each row of $\Gamma $ is
bounded by $k_{n,q}^{1/r}$, an immediate application of Theorem \ref%
{theorem:step1} yields corresponding results for $l_{r}$ ball sparseness.
For example,
\begin{equation*}
\mathbb{P}\left\{ \frac{1}{p}\left\Vert \mathbf{\hat{\Gamma}}-\mathbf{\Gamma
}\right\Vert _{F}^{2}>C_{4}k_{n,q}\left( \frac{\log q}{n}\right)
^{1-r/2}\right\} \leq o\left( p\cdot q^{-\delta _{1}+1}\right) \text{,}
\end{equation*}%
provided that $k_{n,q}=o\left( \frac{n}{\log q}\right) ^{1-r/2}$ and
Conditions 1-2, 3' hold.
\end{remark}

\begin{remark}
\citeasnoun{cai2013covariate} assumes that the matrix $l_{1}$ norm of $\
\left( Cov\left( X^{(1)}\right) \right) ^{-1}$, the inverse of the
covariance matrix of $X^{(1)}$, is bounded, and their tuning parameter
depends on the unknown $l_{1}$ norm. In Theorem \ref{theorem:step1} we don't
need the assumption on the $l_{1}$ norm of \ $\left( Cov\left(
X^{(1)}\right) \right) ^{-1}$ and the tuning parameter $\lambda _{1}$ is
given explicitly.
\end{remark}

To analyze the Step 2 of the procedure in Equation (\ref{Second step: scaled
Lasso}), we need the following assumptions for $\Omega $.

\begin{enumerate}
\item[4.] The precision matrix $\Omega =\left( \omega _{ij}\right) _{p\times
p}$ has the following sparsity condition%
\begin{equation}
\max_{i}\Sigma _{j\neq i}\min \left\{ 1,\frac{\left\vert \omega
_{ij}\right\vert }{\lambda _{2}}\right\} =s_{2}\text{,}
\label{sparsity definition omega}
\end{equation}%
where $\lambda _{2}$ is at an order of $\sqrt{\frac{\log p}{n}}$.

\item[5.] There exists a positive constant $M_{2}$ such that $\left\Vert
\Omega \right\Vert _{l_{\infty }}\leq M_{2}$.
\end{enumerate}

It is convenient to introduce a notation for the covariance matrix of $%
\left( \eta _{i},\eta _{j}\right) ^{T}$ in Equation (\ref{regression}). Let%
\begin{equation*}
\Psi _{A,A}=\Omega _{A,A}^{-1}=\left(
\begin{array}{cc}
\psi _{ii} & \psi _{ij} \\
\psi _{ji} & \psi _{jj}%
\end{array}%
\right) \text{.}
\end{equation*}%
We will estimate $\Psi _{A,A}$ first and show that an efficient estimator of
$\Psi _{A,A}$ yields an efficient estimation of entries of $\Omega _{A,A}$
by inverting the estimator of $\Psi _{A,A}$. Denote a sample version of $%
\Psi _{A,A}$ by
\begin{equation}
\Psi _{A,A}^{ora}=\left( \psi _{kl}^{ora}\right) _{k\in A,l\in A}=\mathbf{%
\epsilon }_{A}^{T}\mathbf{\epsilon }_{A}/n\text{,}  \label{oracle}
\end{equation}%
which is an oracle MLE of $\Psi _{A,A}$, assuming that we know $\mathbf{%
\beta }$, and
\begin{equation}
\Omega _{A,A}^{ora}=\left( \omega _{kl}^{ora}\right) _{k\in A,l\in A}=\left(
\Psi _{A,A}^{ora}\right) ^{-1}\text{.}  \label{oracle1}
\end{equation}%
Let
\begin{equation}
\hat{\Psi}_{A,A}=\mathbf{\hat{\epsilon}}_{A}^{T}\mathbf{\hat{\epsilon}}_{A}/n%
\text{,\ }  \label{estimator}
\end{equation}%
where $\mathbf{\hat{\epsilon}}_{A}$ is defined in Equation (\ref{epsilonhat}%
). Note that $\hat{\Omega}_{A,A}$ defined in Equation (\ref{estimator2}) is
simply the inverse of the estimator $\hat{\Psi}_{A,A}$. The following result
shows that $\hat{\Omega}_{A,A}$ is asymptotically normal and efficient when
both $\Gamma $ and $\Omega $ are sufficient sparse.

\begin{theorem}
\label{theorem:step2} Let $\lambda _{1}$ be defined as in Theorem \ref%
{theorem:step1} with $\delta _{1}\geq 1+\frac{\log p}{\log q}$ and $\lambda
_{2}=\left( 1+\varepsilon _{2}\right) \sqrt{\frac{2\delta _{2}\log p}{n}}$
for any $\delta _{2}\geq 1$ and $\varepsilon _{2}>0$ in Equation (\ref%
{Second step: scaled Lasso}). Assume that
\begin{equation}
s_{1}=o\left( \sqrt{\frac{n}{\log q}}\right) \text{ and }s_{2}=o\left( \sqrt{%
\frac{n}{\log p}}\right) \text{.}  \label{step2: sparsity assumptions}
\end{equation}%
Under Conditions 1-2 and 4-5, and Condition 3 or 3', we have%
\begin{eqnarray}
\mathbb{P}\left\{ \left\Vert \hat{\Psi}_{A,A}-\hat{\Psi}_{A,A}^{ora}\right%
\Vert _{\infty }>C_{5}\left( s_{2}\frac{\log p}{n}+s_{1}\frac{\log q}{n}%
\right) \right\}  &\leq &o\left( p^{-\delta _{2}+1}+pq^{-\delta
_{1}+1}\right) \text{,}  \label{result: step2: close to oracle} \\
\mathbb{P}\left\{ \left\Vert \hat{\Omega}_{A,A}-\Omega
_{A,A}^{ora}\right\Vert _{\infty }>C_{6}\left( s_{2}\frac{\log p}{n}+s_{1}%
\frac{\log q}{n}\right) \right\}  &\leq &o\left( p^{-\delta
_{2}+1}+pq^{-\delta _{1}+1}\right) \text{,}
\label{result: step2: close to oracle omega}
\end{eqnarray}%
for some positive constants $C_{5}$ and $C_{6}$. Furthermore, $\hat{\omega}%
_{ij}$ is asymptotically efficient%
\begin{equation}
\sqrt{nF_{ij}}\left( \hat{\omega}_{ij}-\omega _{ij}\right) \overset{D}{%
\rightarrow }N\left( 0,1\right) \text{,}  \label{efficiency result}
\end{equation}%
when $s_{2}=o\left( \frac{\sqrt{n}}{\log p}\right) $ and $s_{1}=o\left(
\frac{\sqrt{n}}{\log q}\right) $, where%
\begin{equation*}
F_{ij}^{-1}=\omega _{ii}\omega _{jj}+\omega _{ij}^{2}\text{.}
\end{equation*}
\end{theorem}

\begin{remark}
\label{Partial Correlation}The asymptotic normality result can be obtained
for estimation of the partial correlation. Let $r_{ij}=-\omega _{ij}/(\omega
_{ii}\omega _{jj})^{1/2}$ be the partial correlation between $Z_{i}$ and $%
Z_{j}$. Define $\hat{r}_{ij}=-\hat{\omega}_{ij}/(\hat{\omega}_{ii}\hat{\omega%
}_{jj})^{1/2}$. Under the same assumptions in Theorem \ref{theorem:step2},
the estimator $\hat{r}_{ij}$ is asymptotically efficient, i.e., $\sqrt{%
n(1-r_{ij}^{2})^{-2}}(\hat{r}_{ij}-r_{ij})$ $\overset{D}{\rightarrow }%
N\left( 0,1\right) $, when $s_{2}=o\left( \sqrt{n}/\log p\right) $ and $%
s_{1}=o\left( \sqrt{n}/\log q\right) $.
\end{remark}

\begin{remark}
In Equations (\ref{CondiDis2}) and (\ref{regression-data}), we can replace $%
A=\left\{ i,j\right\} $ by a bounded size subset $B\subset \left[ 1:p\right]
$ with cardinality more than $2$. Similar to the analysis of Theorem \ref%
{theorem:step2}, we can show the estimator for any smooth functional of $%
\Omega _{B,B}^{-1}$ is asymptotic normality as shown in %
\citeasnoun{RenSunZhZh2013} for Gaussian graphical model.
\end{remark}

\begin{remark}
\label{finite sample lamda}A stronger result can be obtained for the choice
of $\lambda _{1}$ and $\lambda _{2}$. Theorems \ref{theorem:step1} and \ref%
{theorem:step2} sill hold, when $\lambda _{1}=\left( 1+\varepsilon
_{1}\right) \sqrt{\frac{2\delta _{1}\log \left( q/s_{\max ,1}\right) }{n}}$
and $\lambda _{2}=\left( 1+\varepsilon _{2}\right) \sqrt{\frac{2\delta
_{2}\log \left( p/s_{\max ,2}\right) }{n}}$, where $s_{\max ,1}=o\left(
\frac{\sqrt{n}}{\log q}\right) $ and $s_{\max ,2}=o\left( \frac{\sqrt{n}}{%
\log p}\right) $. Another alternative choice of $\lambda _{1}$ and $\lambda
_{2}$ will be introduced in Section \ref{sec:simu}.
\end{remark}

\section{Adaptive Support Recovery and Estimation of $\Omega $ under Matrix
Norms}

\label{sec: model selection}

In this section, the asymptotic normality result obtained in Theorem \ref%
{theorem:step2} is applied to perform adaptive support recovery and to
obtain rate-optimal estimation of the precision matrix under various matrix $%
l_{w}$ norms. The two-step procedure for support recovery is first removing
the effect of the covariate $X$, then applying ANT (Asymptotically Normal
estimation with Thresholding) procedure. We thus call it ANTAC, which stands
for ANT after Adjusting Covariates.

\subsection{ANTAC for Support Recovery of $\Omega $}

The support recovery on covariate-adjusted Gaussian graphical model has been
studied by several papers, for example, \citeasnoun{yin2013adjusting} and %
\citeasnoun{cai2013covariate}. Denote the support of $\Omega $ by \emph{Supp}%
$\left( \Omega \right) $. In these literature, the theoretical properties on
the support recovery were obtained but they all assumed that $\min_{(i,j)\in
\emph{Supp}\left( \Omega \right) }|\omega _{ij}|\geq CM_{n,p}^{2}\sqrt{\frac{%
\log p}{n}}$, where $M_{n,p}$ is either the matrix $l_{\infty }$ norm or
related to the irrepresentable condition on $\Omega $, which is unknown. The
ANTAC procedure, based on the asymptotic normality estimation in Equation (%
\ref{efficiency result}), performs entry-wise thresholding adaptively to
recover the graph with explicit thresholding levels.

Recall that in Theorem \ref{theorem:step2} we have%
\begin{equation*}
\sqrt{nF_{ij}}\left( \hat{\omega}_{ij}-\omega _{ij}\right) \overset{D}{%
\rightarrow }N\left( 0,1\right) \text{,}
\end{equation*}%
where $F_{ij}=\left( \omega _{ii}\omega _{jj}+\omega _{ij}^{2}\right) ^{-1}$
is the Fisher information of estimating $\omega _{ij}$. Suppose we know this
Fisher information, we can apply a thresholding level $\sqrt{\frac{2\xi
\left( \omega _{ii}\omega _{jj}+\omega _{ij}^{2}\right) \log p}{n}}$ with
any $\xi \geq 2$ for $\hat{\omega}_{ij}$ to correctly distinguish zero and
nonzero entries, noting the total number of edges is $p\left( p-1\right) /2$%
. However, when the variance $\omega _{ii}\omega _{jj}+\omega _{ij}^{2}$ is
unknown, all we need is to plug in a consistent estimator. The ANTAC
procedure is defined as follows
\begin{eqnarray}
\hat{\Omega}_{thr} &=&(\hat{\omega}_{ij}^{thr})_{p\times p},\quad %
\mbox{where~~}\hat{\omega}_{ii}^{thr}=\hat{\omega}_{ii}\text{, and }\hat{%
\omega}_{ij}^{thr}=\hat{\omega}_{ij}1\{|\hat{\omega}_{ij}|\geq \tau _{ij}\}%
\text{ }  \label{adaptive thresholding} \\
\text{with }\tau _{ij} &=&\sqrt{\frac{2\xi _{0}\left( \hat{\omega}_{ii}\hat{%
\omega}_{jj}+\hat{\omega}_{ij}^{2}\right) \log p}{n}}\text{ for }i\neq j%
\text{,}  \label{thresholding level}
\end{eqnarray}%
where $\hat{\omega}_{kl}$ is the consistent estimator of $\omega _{kl}$
defined in (\ref{estimator2}) and $\xi _{0}$ is a tuning parameter which can
be taken as fixed at any $\xi _{0}>2$.

The following sufficient condition for support recovery is assumed in
Theorem \ref{support recovery} below. Define the sign of $\Omega $ by $%
\mathcal{S}(\boldsymbol{\Omega })=\{sgn(\omega _{ij}),~~1\leq i,j\leq p\}$.
Assume that
\begin{equation}
\left\vert \omega _{ij}\right\vert \geq 2\sqrt{\frac{2\xi _{0}\left( \omega
_{ii}\omega _{jj}+\omega _{ij}^{2}\right) \log p}{n}}\text{, }\forall \omega
_{ij}\in \emph{Supp}(\Omega )\text{.}  \label{lower bound for each entry}
\end{equation}%
The following result shows that not only the support of $\Omega $ but also
the signs of the nonzero entries can be recovered exactly by $\hat{\Omega}%
_{thr}$.

\begin{theorem}
\label{support recovery} Assume that Conditions 1-2 and 4-5, and Condition 3
or 3' hold. Let $\lambda _{1}$ be defined as in Theorem \ref{theorem:step1}
with $\delta _{1}\geq 1+\frac{\log p}{\log q}$ and $\lambda _{2}=\left(
1+\varepsilon _{2}\right) \sqrt{\frac{2\delta _{2}\log p}{n}}$ with any $%
\delta _{2}\geq 3$ and $\varepsilon _{2}>0$ in Equation (\ref{Second step:
scaled Lasso}). Also let $\xi _{0}>2$ in the thresholding level (\ref%
{thresholding level}). Under the assumptions (\ref{step2: sparsity
assumptions}) and (\ref{lower bound for each entry}), we have that the ANTAC
defined in (\ref{adaptive thresholding}) recovers the support $\mathcal{S}%
(\Omega )$ consistently, i.e.,
\begin{equation}
\lim_{n\rightarrow \infty }\mathbb{P}\left( \mathcal{S}(\hat{\Omega}_{thr})=%
\mathcal{S}(\Omega )\right) =1\text{.}  \label{result: support recovery}
\end{equation}
\end{theorem}

\begin{remark}
If the assumption (\ref{lower bound for each entry}) does not hold, the
procedure recovers part of the true graph with high partial correlation.
\end{remark}

The proof of Theorem \ref{support recovery} depends on the oracle inequality
(\ref{result: step2: close to oracle omega}) in Theorem \ref{theorem:step2},
a moderate deviation result of the oracle $\hat{\omega}_{ij}$ and a union
bound. The detail of the proof is in spirit the same as that of Theorem 6 in %
\citeasnoun{RenSunZhZh2013}, and thus will be omitted due to the limit of
space.

\subsection{ANTAC for Estimation under the Matrix $l_{w}$ Norm}

In this section, we consider the rate of convergence of a thresholding
estimator of $\Omega $ under the matrix $l_{w}$ norm, including the spectral
norm. The convergence under the spectral norm leads to the consistency of
eigenvalues and eigenvectors estimation. Define $\breve{\Omega}_{thr}$, a
modification of $\hat{\Omega}_{thr}$ defined in (\ref{adaptive thresholding}%
), as follows%
\begin{equation}
\breve{\Omega}_{thr}=(\hat{\omega}_{ij}^{thr}1\left\{ \left\vert \hat{\omega}%
_{ij}\right\vert \leq \log p\right\} )_{p\times p}\text{.}
\label{adaptive threshold- upper bounded}
\end{equation}%
From the idea of the proof of Theorem \ref{support recovery} (see also the
proof of Theorem 6 in \citeasnoun{RenSunZhZh2013}), we see that with high
probability $\left\Vert \hat{\Omega}-\Omega \right\Vert _{\infty }$ is
dominated by $\left\Vert \Omega ^{ora}-\Omega \right\Vert _{\infty
}=O_{p}\left( \sqrt{\frac{\log p}{n}}\right) $ under the sparsity
assumptions (\ref{step2: sparsity assumptions}). The key of the proof in
Theorem\textrm{\ }\ref{matrixnorm} is to derive the upper bound under matrix
$l_{1}$ norm based on the entry-wise supnorm $\left\Vert \breve{\Omega}%
_{thr}-\Omega \right\Vert _{\infty }$. Then the theorem follows immediately
from the inequality $||M||_{l_{w}}\leq ||M||_{l_{1}}$ for any symmetric
matrix $M$ and $1\leq w\leq \infty ,$ which can be proved by applying the
Riesz-Thorin interpolation theorem. The proof follows similarly from that of
Theorem $3$ in \citeasnoun{CZh12Sparse}. We omit the proof due to the limit
of space.

\begin{theorem}
\label{matrixnorm}Assume that Conditions 1-2 and 4-5, and Condition 3 or 3'
hold. Under the assumptions (\ref{step2: sparsity assumptions}) and $n=\max
\left\{ O\left( p^{\xi _{1}}\right) ,O\left( q^{\xi _{2}}\right) \right\} $
with some $\xi _{1},\xi _{2}>0$, the $\breve{\Omega}_{thr}$ defined in (\ref%
{adaptive threshold- upper bounded}) with sufficiently large $\delta _{1}$
and $\delta _{2}$ satisfies, for all $1\leq w\leq \infty $,
\begin{equation}
\mathbb{E}||\breve{\Omega}_{thr}-\Omega ||_{l_{w}}^{2}\leq Cs_{2}^{2}\frac{%
\log p}{n}\text{.}  \label{l0rate}
\end{equation}
\end{theorem}

\begin{remark}
The rate of convergence result in Theorem \ref{matrixnorm} also can be
easily extended to the parameter space in which each row of $\Omega $ is in
a $l_{r}$ ball of radius $k_{n,q}^{1/r}$. See, e.g., Theorem $3$ in %
\citeasnoun{CZh12Sparse}. Under the same assumptions of Theorem\textrm{\ }%
\ref{matrixnorm} except replacing $s_{2}=o\left( \sqrt{n/\log p}\right) $ by
$k_{n,p}^{2}=o\left( \left( n/\log p\right) ^{1-r}\right) $, we have
\begin{equation}
\mathbb{E}||\breve{\Omega}_{thr}-\Omega ||_{l_{w}}^{2}\leq
Ck_{n,p}^{2}\left( \frac{\log p}{n}\right) ^{1-r}\text{.}  \label{lwrate}
\end{equation}
\end{remark}

\begin{remark}
For the Gaussian graphical model without covariate variables, %
\citeasnoun{CLZ12} showed the rates obtained in Equations (\ref{l0rate}) and
(\ref{lwrate}) are optimal when $p\geq cn^{\gamma }$ for some $\gamma >1$
and $k_{n,p}=o\left( n^{1/2}\left( \log p\right) ^{-3/2}\right) $ for the
corresponding parameter spaces of $\Omega $. This implies that our estimator
is rate optimal.
\end{remark}

\section{Simulation studies}

\label{sec:simu}

\subsection{Asymptotic Normal Estimation}

\label{sec:simu estimation}

In this section, we compare the sample distribution of the proposed
estimator for each edge $\omega _{ij}$ with the normal distribution in
Equation (\ref{efficiency result}). Three models are considered with
corresponding $\left\{ p,q,n\right\} $ listed in Table \ref{tab_partial}.
Based on $200$ replicates, the distributions of the estimators match the
asymptotic distributions very well.

Three sparse models are generated in a similar way to those in \citeasnoun%
{cai2013covariate}. The
$p\times q$ coefficient matrix $\Gamma $ is generated as following for all
three models,%
\begin{equation*}
\Gamma _{ij}\overset{i.i.d.}{\sim }N\left( 0,1\right) \cdot \text{Bernoulli}%
(0.025),
\end{equation*}%
where the Bernoulli random variable is independent with the standard
normal variable, taking one with probability $0.025$ and zero otherwise. We
then generate the $p\times p$ precision matrix $\Omega $ with identical
diagonal entries $\omega _{ii}=$ $4$ for the model of $p=200$ or $400$ and $%
\omega _{ii}=$ $5$ for the model of $p=1000$, respectively. The off-diagonal
entries of $\Omega $ are generated i.i.d. as follows for each model,%
\begin{equation*}
\omega _{ij}=\left\{
\begin{array}{cc}
0.3 & \text{with probability }\frac{\pi}{3} \\
0.6 & \text{with probability }\frac{\pi}{3} \\
1 & \text{with probability }\frac{\pi}{3}\\
0 & \text{otherwise}%
\end{array}%
,\right. \text{for } i\neq j
\end{equation*}%
where the probability of being nonzero $\pi=P(\omega _{ij}\neq 0)$ for three models is shown
in Table \ref{tab_partial}. Once both $\Gamma $
and $\Omega $ are chosen for each model, the $n\times p$ outcome matrix $Y$
is simulated from $Y=X\Gamma ^{T}+Z$ where rows of $Z$ are i.i.d. $%
N(0,\Omega ^{-1})$ and rows of $X$ are i.i.d. $N(0,I_{q\times q})$. We
generate $200$ replicates of $X$ and $Y$ for each model.

\begin{table}[!ht]
\caption{Model parameters and simulation results: mean and standard deviation
(in parentheses) of the proposed estimator for the randomly selected entry with value $v_{\omega}$
based on $200$ replicates. }
\label{tab_partial}{\footnotesize \centering
}
\par
{\footnotesize
\begin{tabular}{lccccccc}
&  &  &  &  &  &  &  \\ \hline\hline
($p$, $q$, $n$) & {\scriptsize {$\pi=P(\omega_{ij} \neq 0)$}} & $v_{\omega} = 0$ & $v_{\omega} =
0.3$ & $v_{\omega} = 0.6$ & $v_{\omega} = 1$ &  &  \\ \hline
(200, 100, 400) & 0.025 & -0.015 (0.168) & 0.289 (0.184) & 0.574 (0.165) &
0.986 (0.182) &  &  \\
(400, 100, 400) & 0.010 & -0.003 (0.24) & 0.268 (0.23) & 0.606 (0.23) &
0.954 (0.244) &  &  \\
(1000, 100, 400) & 0.005 & 0.011 (0.21) & 0.292 (0.26) & 0.507 (0.232) &
0.862 (0.236) &  &  \\ \hline\hline
\end{tabular}
}
\par
\begin{flushleft}
\end{flushleft}
\end{table}

We randomly select four entries of $\Omega $ with values $v_{\omega }$ of $0$%
, $0.3$, $0.6$ and $1$ in each model and draw histograms of our estimators
for those four entries based on the 200 replicates. The penalty parameter $%
\lambda _{1}$, which controls the weight of penalty in the regression of the
first step (\ref{first step: scaled lasso}), is set to be $B_{1}/\sqrt{%
n-1+B_{1}^{2}},$ where $B_{1}=q_{t}(1-\frac{1}{2}\left( s_{\max ,1}/q\right)
^{1+\frac{\log p}{\log q}},n-1),$ and $q_{t}(\cdot ,n)$ is the quantile
function of $t$ distribution with degrees of freedom $n$. This parameter $%
\lambda _{1}$ is a finite sample version of the asymptotic level $\sqrt{2(1+%
\frac{\log p}{\log q})\log \left( q/s_{\max ,1}\right) /n}$ we proposed in
Theorem \ref{theorem:step2} and Remark \ref{finite sample lamda}$.$ Here we
pick $s_{\max ,1}=\sqrt{n}/\log q$. The penalty parameter $\lambda _{2}$,
which controls the weight of penalty in the second step (\ref{Second step:
scaled Lasso}), is set to be $B_{2}/\sqrt{n-1+B_{2}^{2}}$ where $%
B_{2}=q_{t}(1-s_{\max ,2}/\left( 2p\right) ,n-1)$, which is asymptotically
equivalent to $\sqrt{2\log \left( p/s_{\max ,2}\right) /n}.$ The $s_{\max
,2} $ is set to be $\sqrt{n}/\log p$.

\begin{figure}[tbp]
\centering
\includegraphics[width=6in]{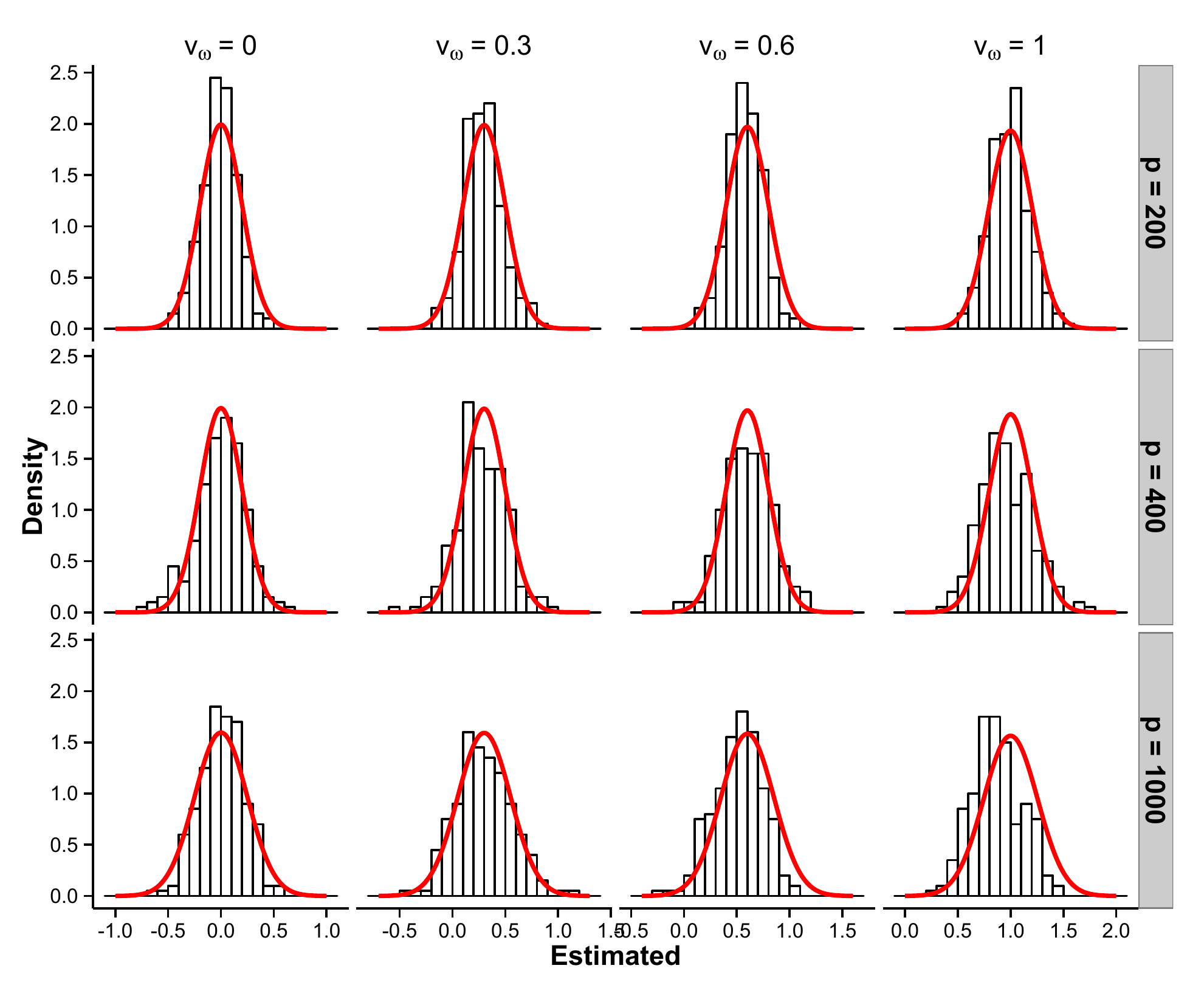}
\caption{The histograms of the estimators for randomly selected entries with values $v_{\omega}=0, 0.3, 0.6$ and $1$ in three models listed in Table \protect\ref{tab_partial}. The
theoretical normal density curves are shown as solid curves. The
variance for each curve is $(\protect\omega_{ii} \protect\omega_{jj} +
\protect\omega_{ij} ^2)/n$, the inverse of the Fisher information.}
\label{hist}
\end{figure}

In Figure \ref{hist}, we show the histograms of the estimators with the theoretical normal density
super-imposed for those randomly selected four entries with values $%
v_{\omega } $ of $0$, $0.3$, $0.6$ and $1$ in each of the three models. The distributions of our estimators
match well with the theoretical normal distributions.

\subsection{Support recovery}

In this section, we evaluate the performance of the proposed ANTAC method
and competing methods in support recovery with different simulation settings.
ANTAC always performs among the best under all model settings. Under the Heterogeneous Model setting, the ANTAC
achieves superior precision and recall rates and performs significantly better than others. Besides, ANTAC is
computationally more efficient compared to a state-of-art method CAPME due to its tuning free property.

\subsubsection{\textbf{Homogeneous Model}}

We consider three models with corresponding $\left\{ p,q,n\right\} $ listed
in Table \ref{tab_model}, which are similar to the models listed in Table %
\ref{tab_model} and used in \cite{cai2013covariate}. Since every model has
identical values along the diagonal, we call them \textquotedblleft
Homogeneous Model\textquotedblright . In terms of the support recovery,
ANTAC performs among the best in all three models, although the performance
from all procedures is not satisfactory due to the intrinsic difficulty of
support recovery problem for models considered.

We generate the $p\times q$ coefficient matrix $\Gamma $ in the same way as
Section \ref{sec:simu estimation},%
\begin{equation*}
\Gamma _{ij}\overset{i.i.d.}{\sim }N\left( 0,1\right) \cdot \text{Bernoulli}%
(0.025).
\end{equation*}%
The off-diagonal entries of the $p\times p$ precision matrix $\Omega $ are
generated as follows,
\begin{equation*}
\omega _{ij}\overset{i.i.d.}{\sim }N\left( 0,1\right) \cdot \text{Bernoulli}%
(\pi ),
\end{equation*}%
where the probability of being nonzero $\pi =P(\omega _{ij}\neq 0)$ is shown
in Table \ref{tab_model} for three models respectively. We generate $50$
replicates of $X$ and $Y$ for each of the three models.

\begin{table}[!ht]
\footnotesize
\centering
\def~{\hphantom{0}}
\caption{Model parameters used in the simulation of support recovery. }
\begin{tabular}{lcccc}
 \\
\hline
\hline
 & ($p$, $q$, $n$) & $P(\Gamma_{ij} \neq 0)$ &  \scriptsize{$\pi=P(\omega_{ij} \neq 0)$, $i \neq j$} \\
\hline
Model 1 & (200, 200, 200) & 0.025 & 0.025 \\
Model 2 & (200, 100, 300) & 0.025 & 0.025 \\
Model 3 & (800, 200, 200) & 0.025 & 0.010 \\
\hline
\hline
\end{tabular}
\begin{flushleft}
\end{flushleft}
\label{tab_model}
\end{table}

We compare our method with graphical Lasso (GLASSO) \cite{friedman2008sparse}%
, a state-of-art method --- CAPME \cite{cai2013covariate} and a conditional
GLASSO procedure (short as cGLASSO), where we apply the same scaled lasso
procedure as the first stage of the proposed method and then estimate the
precision matrix by GLASSO. This cGLASSO procedure is similar to that
considered in \citeasnoun{yin2013adjusting} except that in the first stage %
\citeasnoun{yin2013adjusting} applies ordinary lasso rather than the scaled
lasso, which requires another cross-validation for this step. For GLASSO, the precision matrix is
estimated directly from the sample covariance matrix without taking into
account the effects from $X$. The tuning parameter for the $l_{1}$ penalty
is selected using five-fold cross validation by maximizing the
log-likelihood function. For CAPME, the tuning parameters $\lambda _{1}$ and
$\lambda _{2}$, which control the penalty in the two stages of regression,
are chosen using five-fold cross validation by maximizing the log-likelihood
function. The optimum is achieved via a grid search on $\{(\lambda
_{1},\lambda _{2})\}$. For Models 1 and 2, $10\times 10$ grid is used and for
Model 3, $5\times 5$ grid is used because of the computational burden. Specifically, we use the
CAPME package implemented by the authors of \cite{cai2013covariate}.
For Model 3, each run with $5\times 5$ grid search and five-fold cross validation
takes $160$ CPU hours using one core from PowerEdge M600 nodes $2.33$ GHz and $16-48$ GB RAM,
whereas ANTAC takes $46$ CPU hours. For
ANTAC, the parameter $\lambda _{1}$ is set to be $B_{1}/\sqrt{n-1+B_{1}^{2}}%
, $ where $B_{1}=q_{t}(1-\frac{1}{2}\left( s_{\max ,1}/q\right) ^{1+\frac{%
\log p}{\log q}},n-1)$, $q_{t}(\cdot ,n)$ is the quantile function of $t$
distribution with degrees of freedom $n$ and $s_{\max ,1}=\sqrt{n}/\log q$.
The parameter $\lambda _{2}$, is set to be $B_{2}/\sqrt{n-1+B_{2}^{2}}$
where $B_{2}=q_{t}(1-\left( s_{\max ,2}/p\right) ^{3}/2,n-1)$. For cGLASSO,
the first step is the same as ANTAC. In the second step, the precision matrix
is estimated by applying GLASSO to the estimated $Z$, where the tuning parameter
is selected using five-fold cross validation by maximizing the log-likelihood function.

We evaluate the performance of the estimators for support recovery problem
in terms of the misspecification rate, specificity, sensitivity, precision
and Matthews correlation coefficient, which are defined as,
\begin{eqnarray*}
\text{MISR}(\hat{\Omega},\Omega ) &=&\frac{\text{{\small FN}}+\text{{\small %
FP}}}{p(p-1)}\text{, SPE}=\frac{\text{{\small TN}}}{\text{{\small TN}}+\text{%
{\small FP}}}\text{, SEN}=\frac{\text{{\small TP}}}{\text{{\small TP}}+\text{%
{\small FN}}}\text{,} \\
\text{PRE} &=&\frac{\text{{\small TP}}}{\text{{\small TP}}+\text{{\small FP}}%
}\text{, MCC}=\frac{\text{{\small TP}}\times \text{{\small TN}}-\text{%
{\small FP}}\times \text{{\small FN}}}{\left[ \left( \text{{\small TP}}+%
\text{{\small FP}}\right) \left( \text{{\small TP}}+\text{{\small FN}}%
\right) \left( \text{{\small TN}}+\text{{\small FP}}\right) \left( \text{%
{\small TN}}+\text{{\small FN}}\right) \right] ^{1/2}}\text{.}
\end{eqnarray*}%
Here, {\small TP, TN, FP, FN} are the numbers of true positives, true
negatives, false positives and false negatives respectively. True positives are
defined as the correctly identified nonzero entries of the off-diagonal entries of $\Omega $. For
GLASSO and CAPME, nonzero entries of $\hat{\Omega}$ are selected
as edges with no extra thresholding applied. For ANTAC,
edges are selected by the theoretical bound with $\xi _{0}=2$ . The results
are summarized in Table \ref{tab_support}. It can be seen that ANTAC
achieves superior specificity and precision. Besides, ANTAC has the best
overall performance in terms of the Matthews correlation coefficient.

\begin{figure}[tbp]
\centering
\includegraphics[width=6.5in]{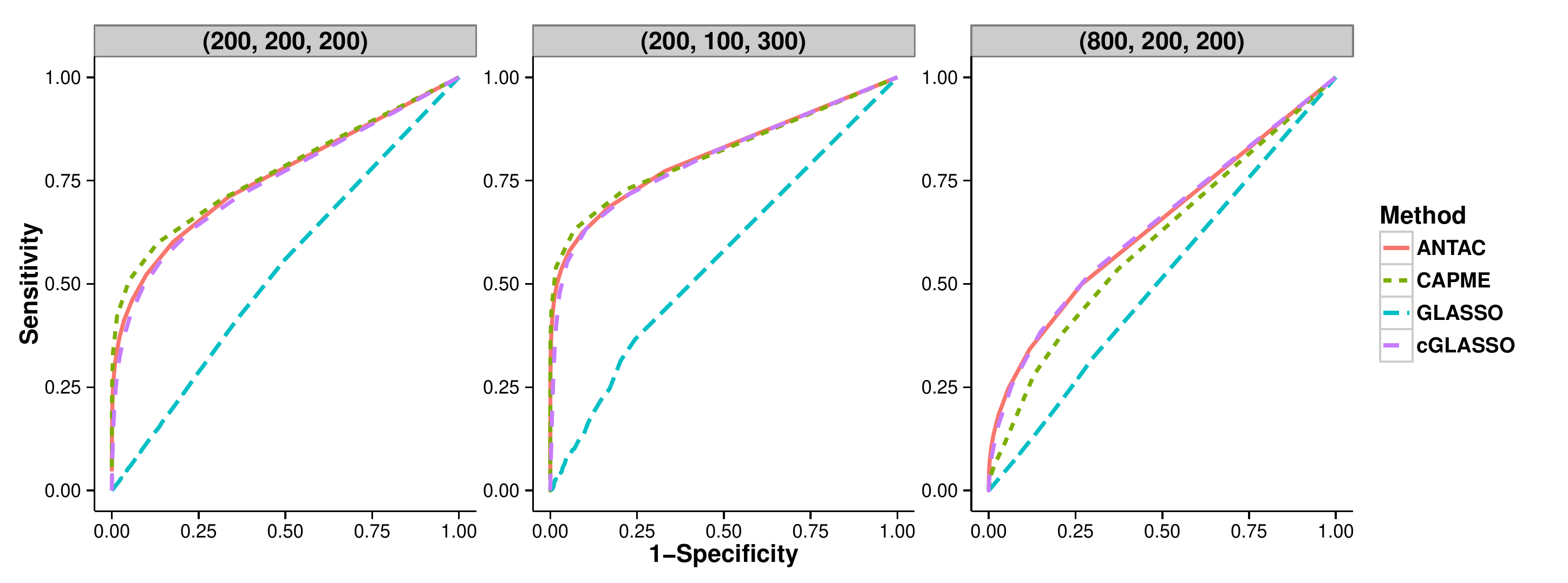}
\caption{The ROC curves for different methods. For GLASSO or cGLASSO, the ROC
curve is obtained by varying its tuning parameter. For CAPME, $%
\protect\lambda_1$ is fixed as the value selected by the cross validation and the ROC
curve is obtained by varying $\lambda_2$.
For ANTAC, the ROC curve is obtained by varying the cut-off on
P-values. }
\label{roc}
\end{figure}

We further construct ROC curves to check how this result would vary by
changing the tuning parameters. For GLASSO or cGLASSO, the ROC curve is
obtained by varying the tuning parameter. For CAPME, $\lambda _{1}$ is fixed
as the value selected by the cross validation and the ROC curve is obtained
by varying $\lambda _{2}$. For proposed ANTAC method, the ROC curve is
obtained by varying the thresholding level $\xi _{0}$. When $p$ is small,
CAPME, cGLASSO and ANTAC have comparable performance. As $p$ grows, both
ANTAC and cGLASSO outperform CAPME.

\begin{table}[!ht]
\scriptsize
\centering
\def~{\hphantom{0}}
\caption{Simulation results of the support recovery for homogeneous models based on 50 replications. Specifically, the performance is measured by misspecification rate, specificity, sensitivity (recall rate), precision and the Matthews correlation coefficient with all the values multiplied by 100. Numbers in parentheses are the simulation standard deviations. }
\begin{tabular}{lcccccccc}
 \\
\hline
\hline
 & ($p$, $q$, $n$)& Method &   MISR & SPE & SEN & PRE & MCC \\
\hline
Model 1 & (200, 200, 200) & GLASSO & 35(1) & 65(1) & 37(2) & 2(0) & 1(1)\\
& & cGLASSO & 25(6) & 76(6) & 64(7) & 6(1) & 8(1)\\
& & CAPME &2(0) & 100(0) & 4(1) & 96(1) & 21(1)\\
& & ANTAC  &2(0) & 100(0) & 4(0) & 88(8) & 18(2)\\
Model 2 & (200, 100, 300) & GLASSO & 43(0) & 57(0) & 51(2) & 3(0) & 3(1)\\
& &cGLASSO & 5(0) & 97(0) & 47(1) & 25(1) & 32(1)\\
& & CAPME & 4(0) & 97(0) & 56(1) & 29(1) & 39(1) \\
& & ANTAC & 2(0) & 100(0) & 22(1) & 97(2) & 46(1)\\
Model 3 & (800, 200, 200)  & GLASSO &  19(1) & 81(1) & 19(1) & 1(0) & 0(0) \\
& & cGLASSO & 1(0) & 100(0) & 0(0) & 100(0) & 2(0)\\
& & CAPME &  1(0) & 100(0) & 0(0) & 0(0) & 0(0)\\
& & ANTAC &  1(0) & 100(0) & 7(0) & 71(2) & 22(1)\\
\hline
\end{tabular}
\begin{flushleft}
\end{flushleft}
\label{tab_support}
\end{table}

The purpose of simulating \textquotedblleft Homogeneous
Model\textquotedblright\ is to compare the performance of ANTAC and other
procedures under models with similar settings used in %
\citeasnoun{cai2013covariate}. Overall the performance from all procedures
is not satisfactory due to the difficulty of support recovery problem. All
nonzero entries are sampled from a standard normal. Hence, most signals are
very weak and hard to be recovered by any method, although ANTAC performs
among the best in all three models.

\subsubsection{\textbf{Heterogeneous Model}}

We consider some models where the diagonal entries of the precision matrix
have different values. These models are different from \textquotedblleft
Homogeneous Model\textquotedblright\ and we call them \textquotedblleft
Heterogeneous Model\textquotedblright . The performance of ANTAC and other
procedures are explored under \textquotedblleft Magnified
Block\textquotedblright\ model and \textquotedblleft Heterogeneous
Product\textquotedblright\ model, respectively. The ANTAC performs
significantly better than GLASSO, cGLASSO and CAMPE in both settings.

\begin{figure}[tbp]
\centering
\includegraphics[width=6in]{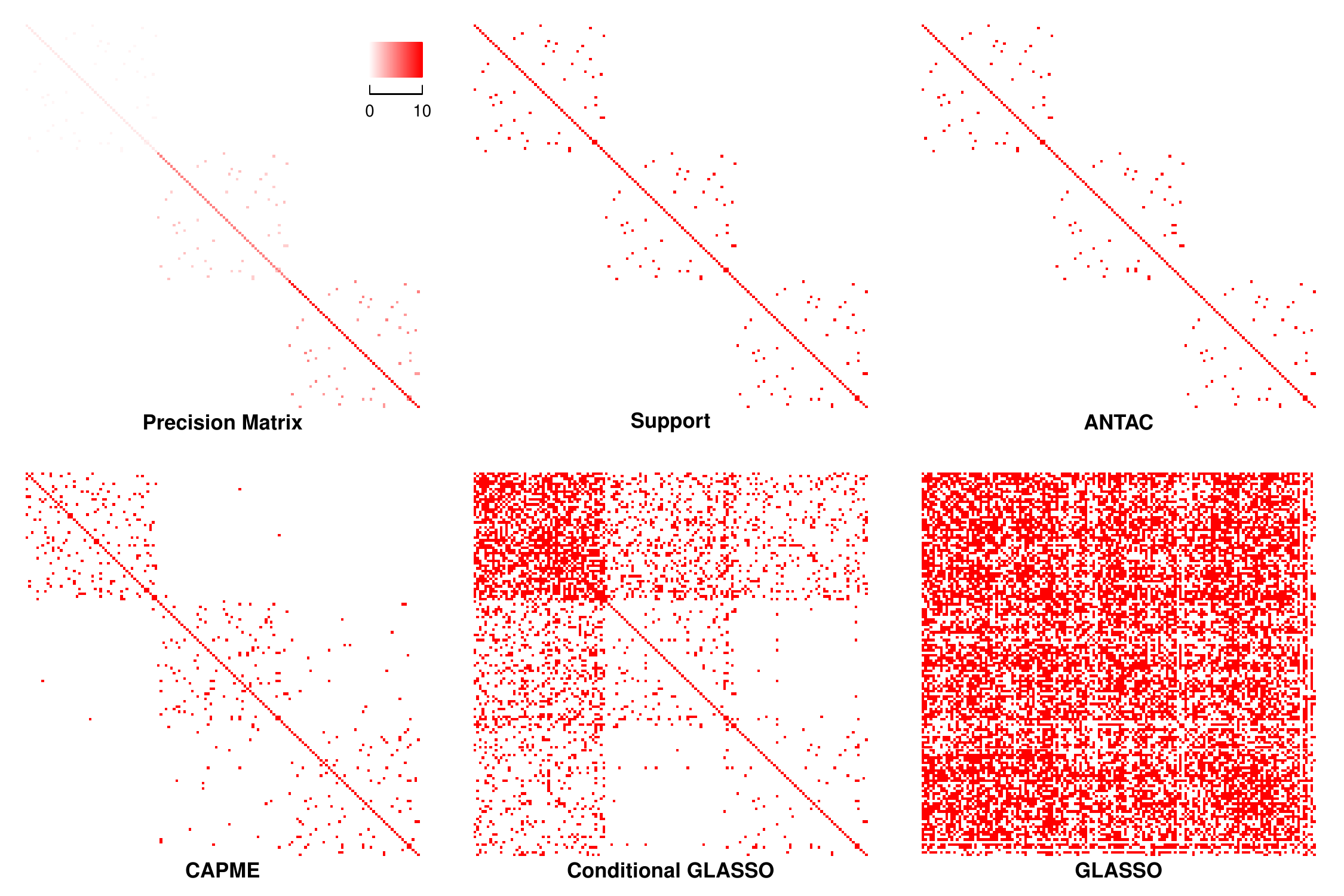}
\caption{Heatmap of support recovery using different methods for a
\textquotedblleft Magnified Block\textquotedblright\ model. }
\label{heatmap1}
\end{figure}

In \textquotedblleft Magnified Block\textquotedblright\ model, we apply the
following randomized procedure to choose $\Omega $ and $\Gamma $. We first
simulate a $50\times 50$ matrix $\Omega _{B}$ with diagonal entries being 1
and each non-diagonal entry i.i.d. being nonzero with $P(\omega _{ij}\neq
0)=0.02$. If $\omega _{ij}\neq 0$, we sample $\omega _{ij}$ from $\{0.4,0.5\}
$. Then we generate two matrices by multiplying $\Omega _{B}$ by $5$ and $10$, respectively. Then we align three matrices along the diagonal, resulting in a
block diagonal matrix $\Omega $ with sequentially magnified signals. A
visualization of the simulated precision matrix is shown in Figure \ref{heatmap1}.
The $150\times 100$ matrix $\Gamma $ is simulated with each entry being
nonzero i.i.d. follows $N(0,1)$ with $P(\Gamma _{ij}\neq 0)=0.05$. Once the
matrices $\Omega $ and $\Gamma $ are chosen, $50$ replicates of $X$ and $Y$
are generated.

\begin{figure}[tbp]
\centering
\includegraphics[width=6in]{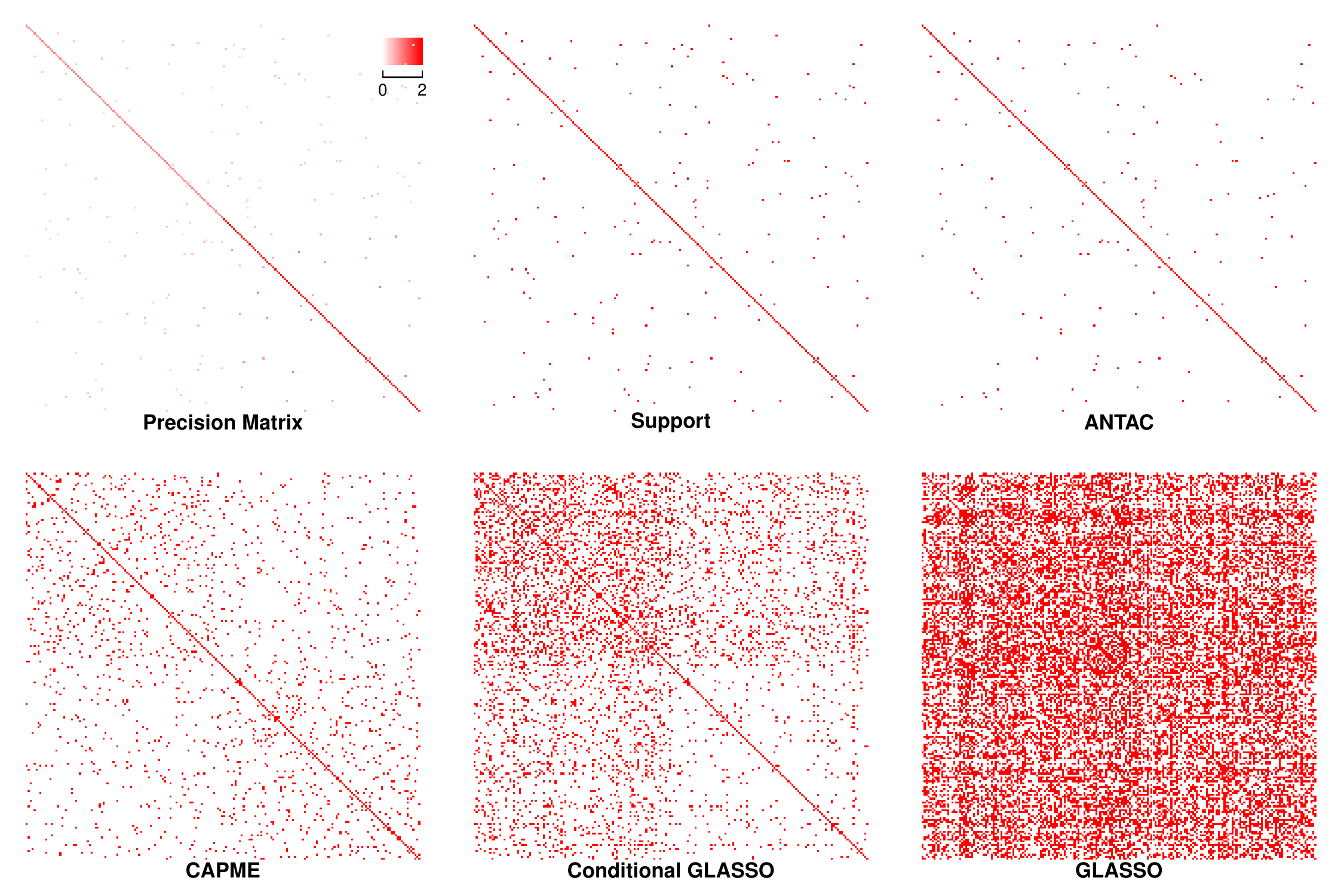}
\caption{Heatmap of support recovery using different methods for a
\textquotedblleft Heterogeneous Product\textquotedblright\ model. }
\label{heatmap2}
\end{figure}

In \textquotedblleft Heterogeneous Product\textquotedblright\ model, the
matrices $\Omega $ and $\Gamma $ are chosen in the following randomized way.
We first simulate a $200\times 200$ matrix $\Omega $ with diagonal entries
being $1$ and each non-diagonal entry i.i.d. being nonzero with $P(\omega
_{ij}\neq 0)=0.005$. If $\omega _{ij}\neq 0$, we sample $\omega _{ij}$ from $%
\{0.4,0.5\}$. Then we replace the $100\times 100$ submatrix at bottom-right
corner by multiplying the $100\times 100$ submatrix at up-left corner by $2$%
, which results in a precision matrix with possibly many different product
values $\omega _{ii}\sigma _{jj}$ over all $i, j$ pairs, where $\sigma _{jj}$
is the $j$th diagonal entry of the covariance matrix $\Sigma =\left( \sigma
_{kl}\right) _{p\times p}=\Omega ^{-1}$. Thus we call it \textquotedblleft
Heterogeneous Product\textquotedblright\ model. A visualization of the
simulated precision matrix is shown in Figure \ref{heatmap2}. The $200\times 100$
matrix $\Gamma $ is simulated with each entry being nonzero i.i.d. follows $%
N(0,1)$ with $P(\Gamma _{ij}\neq 0)=0.05$. Once $\Omega $ and $\Gamma $ are
chosen, $50$ replicates of $X$ and $Y$ are generated.

We compare our method with GLASSO, CAPME and cGLASSO procedures in
\textquotedblleft Heterogeneous Model\textquotedblright . We first compare the
performance of support recovery when a single procedure from each method is
applied. The tuning parameters for each procedure are set in the same way as
in \textquotedblleft Homogeneous Model\textquotedblright\ except that for
CAPME, the optimal tuning parameter is achieved via a $10\times 10$ grid
search on $\{(\lambda _{1},\lambda _{2})\}$ by five-fold cross validation.
We summarize the support recovery results in Table \ref{tab_support_heter}.
A visualization of the support recovery result for a replicate of
\textquotedblleft Magnified Block\textquotedblright\ model and a replicate
of \textquotedblleft Heterogeneous Product\textquotedblright\ model are
shown in Figure \ref{heatmap1} and \ref{heatmap2} respectively. In both
models, ANTAC significantly outperforms others and achieves high precision
and recall rate. Specifically, ANTAC has precision of $0.99$
for two models respectively while no other procedure achieves precision rate
higher than $0.21$ in either model. Besides, ANTAC returns true sparse graph
structure while others report much denser results.

\begin{table}[!ht]
\scriptsize
\centering
\def~{\hphantom{0}}
\caption{Simulation results of the support recovery for heterogeneous models based on 50 replications. The performance is measured by overall error rate, specificity, sensitivity (recall rate), precision and the Matthews correlation coefficient with all the values multiplied by 100. Numbers in parentheses are the simulation standard deviations. }
\begin{tabular}{lcccccccc}
 \\
\hline
\hline
 & ($p$, $q$, $n$)& Method & MISR & SPE & SEN & PRE & MCC \\
\hline
Magnified Block & (150, 100, 300) & GLASSO &54(0) & 46(0) & 80(4) & 1(0) & 4(1)\\
& & cGLASSO &14(0) & 86(0) & 99(1) & 4(0) & 19(0)\\
& & CAPME & 1(0) & 99(0) & 99(0) & 24(2) & 58(1)\\
& & ANTAC  &0(0) & 100(0) & 98(1) & 99(1) & 99(1)\\
Heterogeneous Product & (200, 100, 300) & GLASSO &42(0) & 58(0) & 85(3) & 1(0) & 6(0)\\
& &cGLASSO & 12(0) & 88(0) & 100(0) & 4(0) & 18(0)\\
& & CAPME  & 4(0) & 96(0) & 97(0) & 15(0) & 33(0)\\
& & ANTAC &0(0) & 100(0) & 80(3) & 99(1) & 89(2)\\
\hline
\end{tabular}
\begin{flushleft}
\end{flushleft}
\label{tab_support_heter}
\end{table}

Moreover, we construct the precision-recall curve to compare a sequence of
procedures from different methods. In terms of precision-recall curve, CAPME
has closer performance as the proposed method in \textquotedblleft Magnified
Block\textquotedblright\ model whereas cGLASSO has closer performance as the
proposed method in \textquotedblleft Heterogeneous
Product\textquotedblright\ model, which indicates the proposed method
performs comparable to the better of CAPME and cGLASSO. Another implication
indicates from the precision-recall curve is that while the tuning free ANTAC
method is always close to the best point along the curve created by using
different values of threshold $\xi _{0}$, CAPME and cGLASSO via cross
validation cannot select the optimal parameters on their corresponding
precision-recall curves, even though one of them has potentially good
performance when using appropriate tuning parameters. Here is an explanation
why the ANTAC procedure is better than CAPME and cGLASSO in
\textquotedblleft Magnified Block\textquotedblright\ model and
\textquotedblleft Heterogeneous Product\textquotedblright\ model settings.
Recall that in the second stage, CAPME applies the same penalty level $%
\lambda $ for each entry of the difference $\Omega \hat{\Sigma}-I$, where $%
\hat{\Sigma}$ denotes the sample covariance matrix, but the $i,j$ entry has
variance $\omega _{ii}\sigma _{jj}$ after scaling. Thus CAPME may not
recover the support well in the \textquotedblleft Heterogeneous
Product\textquotedblright\ model settings, where the variances of different
entries may be very different. As for the cGLASSO, we notice that
essentially the same level of penalty is put on each entry $\omega _{ij}$
while the variance of each entry in the $i$th row $\Omega _{i\cdot }$
depends on $\omega _{ii}$. Hence we cannot expect cGLASSO performs very well
in the \textquotedblleft Magnified Block\textquotedblright\ model settings,
where the diagonals $\omega _{ii}$ vary a lot. In contrast, the ANTAC method
adaptively puts the right penalty level (asymptotic variance) for each
estimate of $\omega _{ij}$, therefore it works well in either setting.

\begin{figure}[tbp]
\centering
\includegraphics[width=6in]{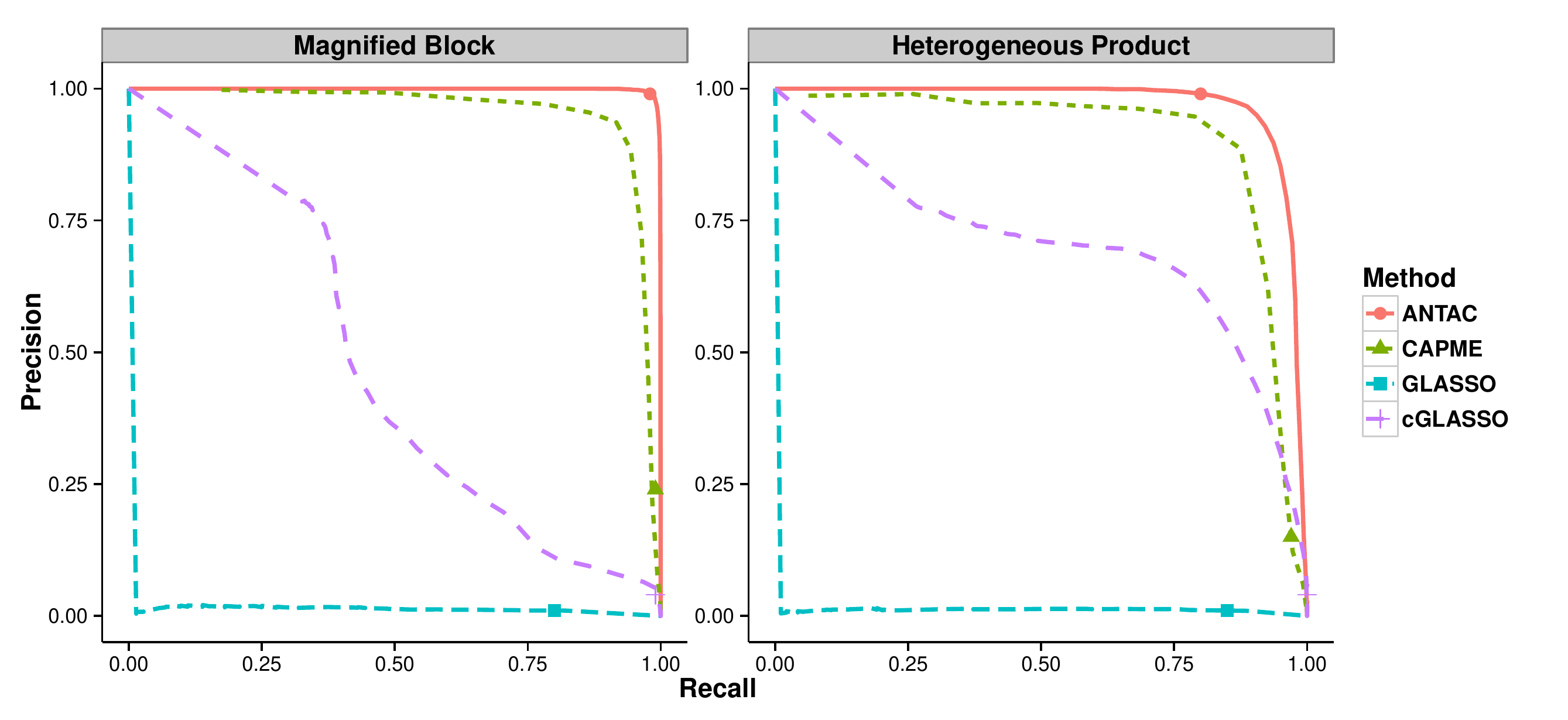}
\caption{The precision-recall curves for \textquotedblleft Magnified Block\textquotedblright\ model and
\textquotedblleft Heterogeneous Product\textquotedblright\ model using different methods. For GLASSO or cGLASSO, the
curve is obtained by varying its tuning parameter. For CAPME, $%
\protect\lambda_1$ is fixed as the value selected by the cross validation and the
curve is obtained by varying $\lambda_2$. For ANTAC, the precision-recall curve is
obtained by varying threshold level $\xi_{0}$. The points on the curves correspond to the results obtained by cross-validation for GLASS, cGLASS and CAPME and by using theoretical threshold level $\xi_{0}=2$ for tuning free ANTAC.}
\label{roc2}
\end{figure}

Overall, the simulation results on heterogeneous models reveal the appealing
practical properties of the ANTAC procedure. Our procedure enjoys tuning
free property and has superior performance. In contrast, it achieves better
precision and recall rate than the results from CAPME and cGLASSO using
cross validation. Although in terms of precision-recall curve, the better of
CAPME and cGLASSO is comparable with our procedure, generally the optimal
sensitivity and specificity could not be obtained through cross-validation.

\section{Application to an eQTL study}

\label{se: real data}

We apply the ANTAC procedure to a yeast dataset from \citeasnoun{smith2008gene}
(GEO accession number GSE9376), which consists of 5,493
gene expression probes and 2,956 genotyped markers measured in 109 segregants derived from a
cross between BY and RM treated with glucose. We find the proposed method
achieves both better interpretability and accuracy in this example.

There are many mechanisms leading to the dependency of genes at the
expression level. Among those, the dependency between transcription factors
(TFs) and their regulated genes has been intensively investigated. Thus the
gene-TF binding information could be utilized as an external biological
evidence to validate and interpret the estimation results. Specifically, we
used the high-confidence TF binding site (TFBS) profiles from m:Explorer, a
database recently compiled using perturbation microarray data, TF-DNA
binding profiles and nucleosome positioning measurements \cite{reimand2012m}.

We first focus our analysis on a medium size dataset that consists of 121 genes on the yeast cell cycle
signaling pathway (from the Kyoto Encyclopedia of Genes and Genomes database
\cite{kanehisa2000kegg}). There are 119 markers marginally associated with at least 3 of
those 121 genes with a Bonferroni corrected P-value less than 0.01.  The parameters $\lambda_1$ and $\lambda_2$
for the ANTAC method are set as described in Theorem \ref%
{support recovery}. 55 edges are identified using a cutoff of 0.01 on the
FDR controlled P-values and 200 edges with a cutoff of 0.05. The number of
edges goes to 375 using a cutoff of 0.1. For the purpose of visualization
and interpretation, we further focus on 55 edges resulted from the cutoff
of 0.01.

\begin{figure}[tbp]
\centering
\includegraphics[width=4in]{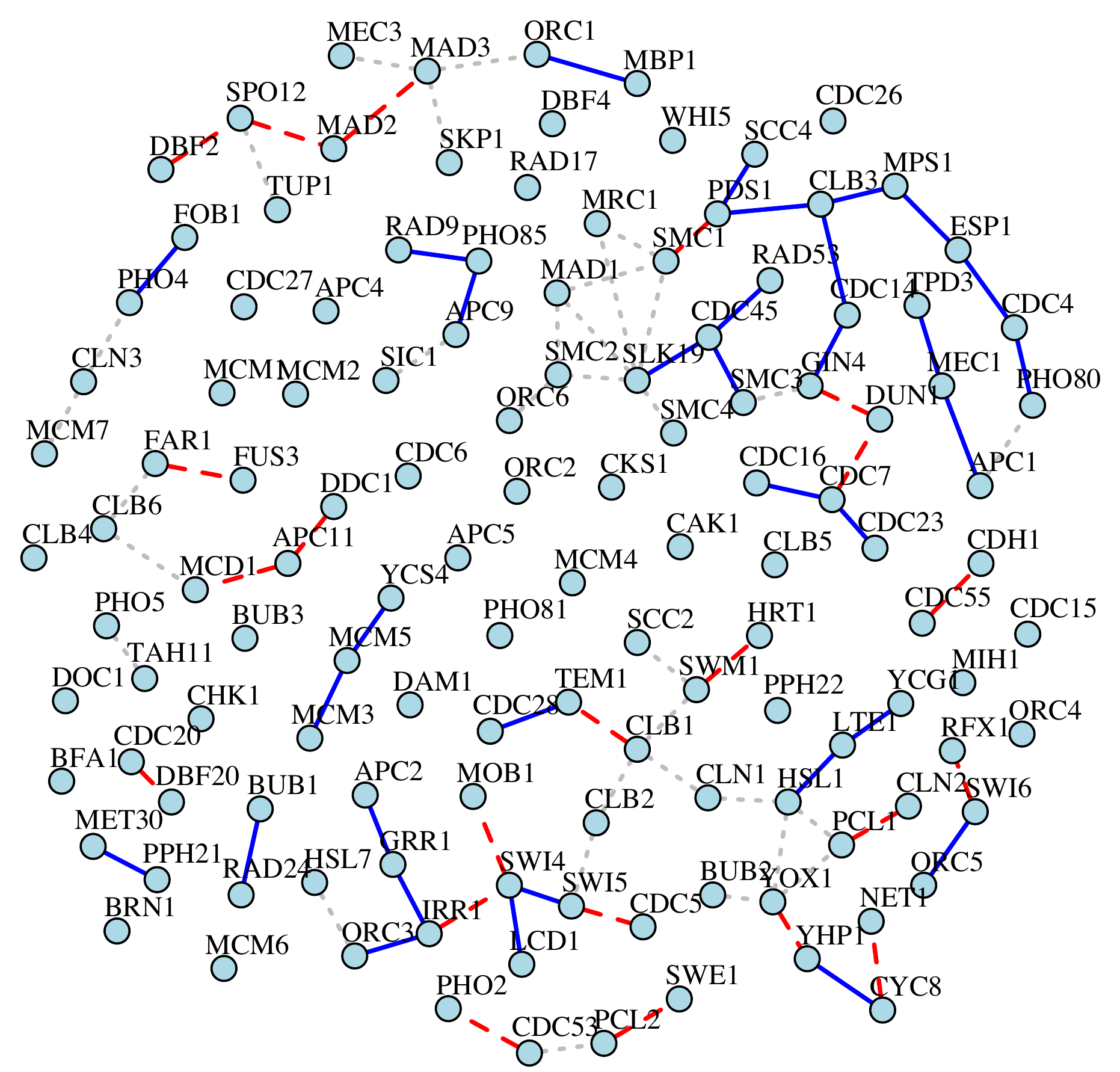}
\caption{Visualization of the network constructed from yeast cell
cycle expression data by CAPME and the proposed ANTAC method.
For ANTAC, 55 edges are identified using a cutoff of 0.01 on the FDR
controlled p-values. For CAPME, top 55 edges are selected based
on the magnitude of partial correlation. 22 common edges detected by both methods
are shown in dashed lines. Edges only detected by the proposed method are shown
in solid lines. CAPME-specific edges are shown in dotted lines. }
\label{cell_cycle}
\end{figure}

We then check how many of these edges involve at least one TF and how many
TF-gene pairs are documented in the m:Explorer database. In 55 detected
edges, 12 edges involve at least one TF and 2 edges are documented. In
addition, we obtain the estimation of precision matrix from CAPME, where
the tuning parameters $\lambda_1=0.356$ and $\lambda_2=0.5$ are chosen by five-fold
cross validation. To compare with the results from ANTAC, we select top
55 edges from CAPME solution based on the magnitude of partial correlation.
Within these 55 edges, 13 edges involve at least one TF and 2 edges are
documented. As shown in Figure \ref{cell_cycle}, 22 edges are detected by
both methods. Our method identifies a promising cell cycle related
subnetwork featured by CDC14, PDS1, ESP1 and DUN1, connecting through GIN4, CLB3 and MPS1.
In the budding yeast, CDC14 is a phosphatase functions essentially in late mitosis. It enables
cells to exit mitosis through dephosphorylation and activation of the
enemies of CDKs \cite{wurzenberger2011phosphatases}. Throughout G1, S/G2 and
early mitosis, CDC14 is inactive. The inactivation is partially achieved by
PDS1 via its inhibition on an activator ESP1 \cite{stegmeier2002separase}.
Moreover, DUN1 is required for the nucleolar localization of CDC14 in DNA
damage-arrested yeast cells \cite{liang2007dna}.

We then extend the analysis to a larger dataset constructed from
GSE9376. For 285 TFs documented in m:Explorer database, expression levels of
20 TFs are measured in GSE9376 with variances greater than 0.25. For these
20 TFs, 875 TF-gene interactions with 377 genes with variances greater than
0.25 are documented in m:Explorer. Applying the screening strategy as the previous example,
we select 644 genetic markers marginally associated
with at least 5 of the 377 genes with a Bonferroni corrected P-value less
than 0.01. We apply the proposed ANTAC method and CAPME to this new dataset.
For ANTAC, the parameters $\lambda_1$ and $\lambda_2$ are set as described in Theorem \ref{support
recovery}. For CAPME, the tuning parameters $\lambda_1=0.089$ and $%
\lambda_2=0.281$ are chosen by five-fold cross validation. We use TF-gene
interactions documented in m:Explorer as an external biological evidence to
validate the results. The results are summarized in Table \ref{tab_real}.
For ANTAC, 540 edges are identified using a cutoff of 0.05 on the FDR
controlled P-values. Within these edges, 67 edges are TF-gene interactions
and 44 out of 67 are documented in m:Explorer. In comparison, 8499 nonzero
edges are detected by CAPME, where 915 edges are TF-gene interactions and
503 out of 915 are documented. This result is hard to interpret
biologically. We further ask if identifying the same number of TF-gene
interactions, which method achieves higher accuracy according to the
concordance with m:Explorer. Based on the magnitude of partial correlation,
we select top 771 edges from the CAPME solution, which capture 67 TF-gene
interactions. Within these interactions, 38 are documented in m:Explorer.
Thus in this example, the proposed ANTAC method achieves both better
interpretability and accuracy.

\begin{table}[!ht]
\footnotesize
\centering
\def~{\hphantom{0}}
\caption{ Results for a dataset consists of 644 markers and 377 genes, which was constructed from GSE9376.}
\begin{tabular}{lccccc}
 \\
\hline
\hline
 Method & Solution  & Total $\#$ of gene-gene & $\#$ of TF-gene & $\#$ of TF-gene & Documented  \\
 & Criteria & interactions & interactions & interactions documented & Proportion\\
\hline
ANTAC & FDR controlled P-values $\leq$0.05  & 540 & 67 & 44 &  65\%\\
CAPME & Magnitude of partial correlation & 771 & 67 & 38 & 57\%\\
CAPME & Nonzero entries & 8499 & 915 & 503 & 55\% \\
\hline
\hline
\end{tabular}
\begin{flushleft}
\end{flushleft}
\label{tab_real}
\end{table}

\section{Proof of Main Theorems}

\label{se: proof of main theorems}

In this section, we will prove the main results Theorems \ref{theorem:step1}
and \ref{theorem:step2}.

\subsection{Proof of Theorem \protect\ref{theorem:step1}}

\label{se: proof of theorem 1}

This proof is based on the key lemma, Lemma \ref{KeyLemma:Scaled Lasso},
which is deterministic in nature. We apply Lemma \ref{KeyLemma:Scaled Lasso}
with $\mathbf{R=D}b^{true}+\mathbf{E}$ replaced by $\mathbf{Y}_{j}\mathbf{=X}%
\gamma _{j}+\mathbf{Z}_{j},$ $\lambda $ replaced by $\lambda _{1}$ and
sparsity $s$ replaced by $s_{1}$. The following lemma is the key to the
proof.

\begin{lemma}
\label{lemma: step1_ prob} There exist some constants $C_{k}^{\prime }$, $%
1\leq k\leq 3$ such that for each $1\leq j\leq p$,
\begin{eqnarray}
\left\Vert \mathbf{\gamma }_{j}-\mathbf{\hat{\gamma}}_{j}\right\Vert _{1}
&\leq &C_{1}^{\prime }\lambda _{1}s_{1}\text{,}  \label{LemmaInequ2} \\
\left\Vert \mathbf{\gamma }_{j}-\mathbf{\hat{\gamma}}_{j}\right\Vert  &\leq
&C_{2}^{\prime }\lambda _{1}\sqrt{s_{1}}\text{,}  \label{LemmaInequ3} \\
\left\Vert \mathbf{X}\left( \mathbf{\gamma }_{j}-\mathbf{\hat{\gamma}}%
_{j}\right) \right\Vert ^{2}/n &\leq &C_{3}^{\prime }\lambda _{1}^{2}s_{1}%
\text{,}  \label{LemmaInequ5}
\end{eqnarray}%
with probability $1-o\left( q^{-\delta _{1}+1}\right) $.
\end{lemma}

With the help of the lemma above, it's trivial to finish our proof. In fact,
$\left\Vert \mathbf{\hat{Z}}_{j}-\mathbf{Z}_{j}\right\Vert ^{2}=\left\Vert
\mathbf{X}\left( \mathbf{\gamma }_{j}-\mathbf{\hat{\gamma}}_{j}\right)
\right\Vert ^{2}$ and hence Equation (\ref{First step: z bound}) immediately
follows from result (\ref{LemmaInequ5}). Equations (\ref{First step: gamma
Lsup bound}) and (\ref{First step: gamma Fro bound}) are obtained by the
union bound and Equations (\ref{LemmaInequ2}) and (\ref{LemmaInequ3})
because of the following relationship,%
\begin{eqnarray*}
\left\Vert \mathbf{\hat{\Gamma}}-\mathbf{\Gamma }\right\Vert _{l_{\infty }}
&=&\max_{j}\left\Vert \mathbf{\gamma }_{j}-\mathbf{\hat{\gamma}}%
_{j}\right\Vert _{1}\text{,} \\
\frac{1}{p}\left\Vert \mathbf{\hat{\Gamma}}-\mathbf{\Gamma }\right\Vert
_{F}^{2} &=&\frac{1}{p}\sum_{j=1}^{p}\left\Vert \mathbf{\gamma }_{j}-\mathbf{%
\hat{\gamma}}_{j}\right\Vert ^{2}\text{.}
\end{eqnarray*}%
It is then enough to prove Lemma \ref{lemma: step1_ prob} to complete the
proof of Theorem \ref{theorem:step1}.

\subsubsection{\textbf{Proof of Lemma \protect\ref{lemma: step1_ prob}}}

The Lemma is an immediate consequence of Lemma \ref{KeyLemma:Scaled Lasso}
applied to $\mathbf{Y}_{j}\mathbf{=X}\gamma _{j}+\mathbf{Z}_{j}$ with tuning
parameter $\lambda _{1}$ and sparsity $s_{1}$. To show the union of
Equations (\ref{LemmaInequ2})-(\ref{LemmaInequ5}) holds with high
probability $1-o\left( q^{-\delta _{1}+1}\right) $, we only need to check
the following conditions of the Lemma \ref{KeyLemma:Scaled Lasso} hold with
probability at least $1-o\left( q^{-\delta _{1}+1}\right) $,%
\begin{eqnarray*}
I_{1} &=&\left\{ \nu \leq \theta ^{ora}\lambda _{1}\frac{\xi -1}{\xi +1}%
\left( 1-\tau \right) \text{ for some }\xi >1\right\} \text{,} \\
I_{2} &=&\left\{ \frac{\left\Vert \mathbf{X}_{k}\right\Vert }{\sqrt{n}}\in %
\left[ 1/A_{1},A_{1}\right] \text{ for all }k\right\} \text{,} \\
I_{3} &=&\left\{ \theta ^{ora}\in \left[ 1/A_{2},A_{2}\right] \text{, where }%
\theta ^{ora}=\frac{\left\Vert \mathbf{Z}_{j}\right\Vert }{\sqrt{n}}\text{.}%
\right\} \text{,} \\
I_{4} &=&\left\{
\begin{array}{c}
\frac{\mathbf{W}^{T}\mathbf{W}}{n}\text{ satisfies lower-RE with }\left(
\alpha _{1},\zeta \left( n,q\right) \right) \text{ s.t.} \\
s_{1}\zeta \left( n,q\right) 8\left( 1+\xi \right) ^{2}A_{1}^{2}\leq \min \{%
\frac{\alpha _{1}}{2},1\}%
\end{array}%
\right\} \text{,}
\end{eqnarray*}%
where we set $\xi =3/\varepsilon _{1}+1$ for the current setting, $%
A_{1}=C_{A_{1}}\max \left\{ B,\sqrt{M_{1}}\right\} $ under Condition $3$ and
$A_{1}=C_{A_{1}}\sqrt{M_{1}}$ under Condition $3^{\prime }$ for some
universal constant $C_{A_{1}}>0$, $A_{2}=\sqrt{2M_{2}}$, $\alpha _{1}=\frac{1%
}{2M_{1}}$ and $\zeta \left( n,q\right) =o(1/s_{1})$. Let us still define $%
\mathbf{W=X}\cdot diag\left( \frac{\sqrt{n}}{\left\Vert \mathbf{X}%
_{k}\right\Vert }\right) $ as the standardized $\mathbf{X}$ in Lemma \ref%
{KeyLemma:Scaled Lasso} of Section \ref{se: key lemma}.

We will show that%
\begin{eqnarray*}
\mathbb{P}\left\{ I_{1}^{c}\right\}  &\leq &O\left( q^{-\delta _{1}+1}/\sqrt{%
\log q}\right) \text{,} \\
\mathbb{P}\left\{ I_{i}^{c}\right\}  &\leq &o(q^{-\delta _{1}})\text{ for }%
i=2,3\text{ and }4\text{,}
\end{eqnarray*}%
which implies
\begin{equation*}
\mathbb{P}\left\{ E_{j}\right\} \geq 1-o\left( q^{-\delta _{1}+1}\right)
\text{,}
\end{equation*}%
where $E_{j}$ is the union of $I_{1}$ to $I_{4}$. We will first consider $%
\mathbb{P}\left\{ I_{2}^{c}\right\} $ and $\mathbb{P}\left\{
I_{3}^{c}\right\} $, then $\mathbb{P}\left\{ I_{4}^{c}\right\} $, and leave $%
\mathbb{P}\left\{ I_{1}^{c}\right\} $ to the last, which relies on the
bounds for $\mathbb{P}\left\{ I_{i}^{c}\right\} $, $2\leq i\leq 4$.

\textbf{(1).} To study $\mathbb{P}\left\{ I_{2}^{c}\right\} $ and $\mathbb{P}%
\left\{ I_{3}^{c}\right\} $, we need the following Bernstein-type inequality
(See e.g. \cite{vershynin2010introduction}, Section $5.2.4$ for the
inequality and definitions of norms $\left\Vert \cdot \right\Vert _{\varphi
_{2}}$ and $\left\Vert \cdot \right\Vert _{\varphi _{1}}$) to control the
tail bound for sum of i.i.d. sub-exponential variables,
\begin{equation}
\mathbb{P}\left\{ \left\vert \sum_{i=1}^{n}U_{i}\right\vert \geq tn\right\}
\leq 2\exp \left( -cn\min \left\{ \frac{t^{2}}{K^{2}},\frac{t}{K}\right\}
\right) \text{,}  \label{Bernstein}
\end{equation}%
for $t>0$, where $U_{i}$ are i.i.d. centered sub-exponential variables with
parameter $\left\Vert U_{i}\right\Vert _{\varphi _{1}}\leq K$ and $c$ is
some universal constant. Notice that $\theta ^{ora}=\left\Vert \mathbf{Z}%
_{j}\right\Vert /\sqrt{n}$ with $\mathbf{Z}_{j}^{(1)}\sim N\left( 0,\sigma
_{jj}\right) $, and $\mathbf{X}_{k}^{(1)}$ is sub-gaussian with parameter $%
\left\Vert \mathbf{X}_{k}^{(1)}\right\Vert _{\varphi _{2}}\in \left[
c_{\varphi 2}M_{1}^{-1/2},C_{\varphi 2}B\right] $ by Conditions $2$ and $3$
with some universal constants $c_{\varphi 2}$ and $C_{\varphi 2}$ (all
formulas involving $\varphi _{1}$ or $\varphi _{2}$ parameter of $\mathbf{X}%
_{k}^{(1)}$ replace \textquotedblleft $B$\textquotedblright\ by
\textquotedblleft $\sqrt{M_{1}}$\textquotedblright\ under Condition $%
3^{\prime }$ hereafter). The fact that sub-exponential is sub-gaussian
squared implies that there exists some universal constant $C_{1}>0$ such
that
\begin{eqnarray*}
&&n\left( \left( \theta ^{ora}\right) ^{2}-\sigma _{jj}\right) \text{ is sum
of i.i.d. sub-exponential with }\varphi _{1}\text{ parameter }C_{1}\sigma
_{jj}\text{,} \\
&&\left\Vert \mathbf{X}_{k}\right\Vert ^{2}-n\mathbb{E}x_{k}^{2}\text{ is
sum of i.i.d. sub-exponential with }\varphi _{1}\text{\ parameter }%
\left\Vert \left( \mathbf{X}_{k}^{(1)}\right) ^{2}\right\Vert _{\varphi _{1}}%
\text{.}
\end{eqnarray*}%
Note that $\sigma _{jj}\in \left[ 1/M_{2},M_{2}\right] $ by Condition $2$
and $\left\Vert \left( \mathbf{X}_{k}^{(1)}\right) ^{2}\right\Vert _{\varphi
_{1}}\in \left[ c_{\varphi 1}^{\prime }M_{1}^{-1},C_{\varphi 1}^{\prime
}B^{2}\right] $ with some universal constants $c_{\varphi 1}^{\prime }$ and $%
C_{\varphi 1}^{\prime }$. Let $A_{1}=C_{A_{1}}\max \left\{ B,\sqrt{M_{1}}%
\right\} $ and $A_{2}=\sqrt{2M_{2}}$ for some large constant $C_{A_{1}}$.
Equation (\ref{Bernstein}) with a sufficiently small constant $c_{0}$
implies
\begin{eqnarray}
\mathbb{P}\left\{ I_{3}^{c}\right\}  &=&\mathbb{P}\left\{ \left( \theta
^{ora}\right) ^{2}\notin \left[ 1/A_{2}^{2},A_{2}^{2}\right] \right\} \leq
\mathbb{P}\left\{ n\left\vert \left( \theta ^{ora}\right) ^{2}-\sigma
_{jj}\right\vert \geq \frac{\sigma _{jj}}{2}n\right\}   \notag \\
&\leq &2\exp \left( -C^{^{\prime }}n\right) \leq o(q^{-\delta _{1}})\text{,}
\label{bound for I3}
\end{eqnarray}%
and
\begin{eqnarray}
\mathbb{P}\left\{ I_{2}^{c}\right\}  &=&\mathbb{P}\left\{ \frac{\left\Vert
\mathbf{X}_{k}\right\Vert ^{2}}{n}\notin \left[ 1/A_{1}^{2},A_{1}^{2}\right]
\text{ for some }k\right\}   \notag \\
&\leq &q\mathbb{P}\left\{ \left\vert \left\Vert \mathbf{X}_{k}\right\Vert
^{2}-n\mathbb{E}x_{k}^{2}\right\vert >c_{0}\left\Vert \left( \mathbf{X}%
_{k}^{(1)}\right) ^{2}\right\Vert _{\varphi _{1}}n\right\} \leq q2\exp
\left( -C^{\prime }n\right) \leq o(q^{-\delta _{1}})\text{.}
\label{bound for  I2}
\end{eqnarray}

\textbf{(2)}. To study the lower-RE condition of $\frac{\mathbf{W}^{T}%
\mathbf{W}}{n}$, we essentially need to study $\frac{\mathbf{X}^{T}\mathbf{X}%
}{n}$. On the event $I_{2}$, it's easy to see that if $\frac{\mathbf{X}^{T}%
\mathbf{X}}{n}$ satisfies lower-RE with $\left( \alpha _{x},\zeta _{x}\left(
n,q\right) \right) $, then $\frac{\mathbf{W}^{T}\mathbf{W}}{n}$ satisfies
lower-RE with $\left( \alpha _{1},\zeta \right) =\left( \alpha
_{x}A_{1}^{-2},\zeta _{x}A_{1}^{2}\right) $, since $\mathbf{W}=\mathbf{X}%
diag\left( \frac{\sqrt{n}}{\left\Vert X_{k}\right\Vert }\right) $ and $\frac{%
\left\Vert \mathbf{X}_{k}\right\Vert }{\sqrt{n}}\in \left[ 1/A_{1},A_{1}%
\right] $ on event $I_{2}$. Moreover, to study the lower-RE condition, the
following lemma implies that we only need to consider the behavior of $\frac{%
\mathbf{X}^{T}\mathbf{X}}{n}$ on sparse vectors.

\begin{lemma}
\label{lower RE Lemma}For any symmetric matrix $\Delta _{q\times q}$,
suppose $\left\vert v^{T}\Delta v\right\vert \leq \delta $ for any unit $2s$
sparse vector $v\in \mathbb{R}^{q}$, i.e. $\left\Vert v\right\Vert =1$ and $%
v\in \mathbb{B}_{0}\left( 2s\right) =\left\{ a:\sum_{i=1}^{q}1\left\{
a_{i}\neq 0\right\} \leq 2s\right\} $, then we have%
\begin{equation*}
\left\vert v^{T}\Delta v\right\vert \leq 27\delta \left( \left\Vert
v\right\Vert ^{2}+\frac{1}{s}\left\Vert v\right\Vert _{1}^{2}\right) \text{
for any }v\in \mathbb{R}^{q}\text{.}
\end{equation*}
\end{lemma}

See Supplementary Lemma $12$ in \cite{loh2012high} for the proof. With a
slight abuse of notation, we define $\Sigma _{x}=Cov\left( X^{\left(
1\right) }\right) $. By applying Lemma \ref{lower RE Lemma} on $\frac{%
\mathbf{X}^{T}\mathbf{X}}{n}-\Sigma _{x}$ and Condition $2$, $v^{T}\Sigma
_{x}v\geq $\ $\frac{1}{M_{1}}\left\Vert v\right\Vert ^{2}$, we know that $%
\frac{\mathbf{X}^{T}\mathbf{X}}{n}$ satisfies lower-RE $\left( \alpha
_{x},\zeta _{x}\left( n,q\right) \right) $ with
\begin{equation}
\alpha _{x}\geq \frac{1}{M_{1}}\left( 1-\frac{27}{L}\right) \text{ and }%
\zeta _{x}\left( n,q\right) \leq \frac{27}{s_{1}M_{1}L}\text{,}
\label{RE parameters of x}
\end{equation}%
provided
\begin{equation}
\left\vert v^{T}\left( \frac{\mathbf{X}^{T}\mathbf{X}}{n}-\Sigma _{x}\right)
v\right\vert \leq \frac{1}{LM_{1}}\text{ for all }v\in \mathbb{B}_{0}\left(
2s_{1}\right) \text{ with }\left\Vert v\right\Vert =1\text{,}
\label{RE key fact}
\end{equation}%
which implies that the population covariance matrix $\Sigma _{x}$ and its
sample version $\frac{\mathbf{X}^{T}\mathbf{X}}{n}$ behave similarly on all $%
2s_{1}$ sparse vectors.

Now we show Equation (\ref{RE key fact}) holds under Conditions $3$ and $%
3^{\prime }$ respectively for a sufficiently large constant $L$ such that
the inequality in the event $I_{4}$ holds. Under Condition $3^{\prime }$, $%
X^{\left( 1\right) }$ is jointly sub-gaussian with parameter $\left(
M_{1}\right) ^{1/2}$. A routine one-step chaining (or $\delta $-net)
argument implies that there exists some constant $c_{sg}>0$ such that
\begin{equation}
\mathbb{P}\left\{ \sup_{v\in \mathbb{B}_{0}\left( 2s_{1}\right) }\left\vert
v^{T}\left( \frac{\mathbf{X}^{T}\mathbf{X}}{n}-\Sigma _{x}\right)
v\right\vert >t\left\Vert v\right\Vert ^{2}\right\} \leq 2\exp \left(
-c_{sg}n\min \left\{ \frac{t^{2}}{M_{1}^{2}},\frac{t}{M_{1}}\right\}
+2s_{1}\log q\right) \text{.}  \label{RE for subgaussian}
\end{equation}%
See e.g. Supplementary Lemma $15$ in \cite{loh2012high} for the proof. Hence
by picking small $t=\frac{1}{LM_{1}}$ with any fixed but arbitrary large $L,$
the sparsity condition (\ref{sparsity condition s1}) $s_{1}=o\left( \frac{n}{%
\log q}\right) $ and Equation (\ref{RE for subgaussian}) imply that Equation
(\ref{RE key fact}) holds with probability $1-$ $o\left( p^{-\delta
_{1}}\right) $.

Under Condition $3$, if $s_{1}=o\left( \sqrt{\frac{n}{\log q}}\right) $,
Hoeffding's inequality and a union bound imply that
\begin{equation*}
\mathbb{P}\left\{ \left\Vert \frac{\mathbf{X}^{T}\mathbf{X}}{n}-\Sigma
_{x}\right\Vert _{\infty }>2B\sqrt{\frac{(1+\delta _{1})\log q}{n}}\right\}
\leq 2q^{-2\delta _{1}}\text{,}
\end{equation*}%
where the norm $\left\Vert \cdot \right\Vert _{\infty }$ denotes the
entry-wise supnorm and $\left\Vert X^{\left( 1\right) }\right\Vert _{\infty
}\leq B$ (see, e.g. \cite{massart2007concentration} Proposition $2.7$). Thus
with probability $1-$ $o\left( q^{-\delta _{1}}\right) $, we have for any $%
v\in \mathbb{B}_{0}\left( 2s_{1}\right) $ with $\left\Vert v\right\Vert =1$,
\begin{equation*}
\left\vert v^{T}\left( \frac{\mathbf{X}^{T}\mathbf{X}}{n}-\Sigma _{x}\right)
v\right\vert \leq \left\Vert \frac{\mathbf{X}^{T}\mathbf{X}}{n}-\Sigma
_{x}\right\Vert _{\infty }\left\Vert v\right\Vert _{1}^{2}\leq 2B\sqrt{\frac{%
(1+\delta _{1})\log q}{n}}\frac{1}{2s_{1}}=o(1)\text{,}
\end{equation*}%
where the last inequality follows from $\left\Vert v\right\Vert _{1}\leq
\sqrt{2s_{1}}\left\Vert v\right\Vert $ for any $v\in \mathbb{B}_{0}\left(
2s_{1}\right) $. Therefore we have Equation (\ref{RE key fact}) holds with
probability $1-$ $o\left( q^{-\delta _{1}}\right) $ for any arbitrary large $%
L$. If $s_{1}=o\left( \frac{n}{\log ^{3}n\log q}\right) $, an involved
argument using Dudley's entropy integral and Talagrand's concentration
theorem for empirical processes implies (see, \cite%
{rudelson2011reconstruction} Theorem $23$ and its proof) Equation (\ref{RE
key fact}) holds with probability $1-$ $o\left( q^{-\delta _{1}}\right) $
for any fixed but arbitrary large $L$.

Therefore we showed that under Condition $3$ or $3^{\prime }$, Equation (\ref%
{RE key fact}) holds with probability $1-$ $o\left( q^{-\delta _{1}}\right) $
for any arbitrary large $L.$ Consequently, Equation (\ref{RE parameters of x}%
) with $\left( \alpha _{1},\zeta \right) =\left( \alpha _{x}A_{1}^{-2},\zeta
_{x}A_{1}^{2}\right) $ on event $I_{2}$ implies that we can pick $\alpha
_{1}=\frac{1}{2M_{1}}$ and sufficiently small $\zeta $ such that event $I_{4}
$ holds with probability $1-$ $o\left( q^{-\delta _{1}}\right) $.

\textbf{(3).} Finally we study the probability of event $I_{1}$. The
following tail probability of $t$ distribution is helpful in the analysis.

\begin{proposition}
\label{t distribution} Let $T_{n}$ follows a $t$ distribution with $n$
degrees of freedom. Then there exists $\epsilon _{n}\rightarrow 0$ as $%
n\rightarrow \infty $ such that $\forall t>0$%
\begin{equation}
\mathbb{P}\left\{ T_{n}^{2}>n\left( e^{2t^{2}/(n-1)}-1\right) \right\} \leq
\left( 1+\epsilon _{n}\right) e^{-t^{2}}/\left( \pi ^{1/2}t\right) \text{.}
\notag
\end{equation}
\end{proposition}

Please refer to \cite{SZH12-ScaledLaoo} Lemma 1 for the proof. Recall that $%
I_{1}=\left\{ \frac{\nu }{\theta ^{ora}}\leq \lambda _{1}\frac{\xi -1}{\xi +1%
}\left( 1-\tau \right) \right\} $, where $\tau $ defined in Equation (\ref%
{tau}) satisfies that $\tau =O\left( s_{1}\lambda _{1}^{2}\right) =o\left(
1\right) $ on $\cap _{i=2}^{4}I_{i}$. From the definition of $\nu $ in
Equation (\ref{v}) we have
\begin{equation*}
\frac{\nu }{\theta ^{ora}}=\max_{k}\left\vert h_{k}\right\vert \text{, with }%
h_{k}=\frac{\mathbf{W}_{k}^{T}\mathbf{Z}_{j}}{n\theta ^{ora}}\text{,}
\end{equation*}%
where each column of $\mathbf{W}$ has norm $\left\Vert \mathbf{W}%
_{k}\right\Vert =\sqrt{n}$. Given $\mathbf{X}$, equivalently $\mathbf{W}$,
it's not hard to check that we have $\frac{\sqrt{n-1}h_{k}}{\sqrt{1-h_{k}^{2}%
}}\sim t_{\left( n-1\right) }$ by the normality of $\mathbf{Z}_{j}$, where $%
t_{\left( n-1\right) }$ is the $t$ distribution with $\left( n-1\right) $
degrees of freedom. From Proposition \ref{t distribution} we have%
\begin{eqnarray*}
&&\mathbb{P}\left\{ \left\vert h_{k}\right\vert >\sqrt{\frac{2t^{2}}{n}}%
\right\}  \\
&=&\mathbb{P}\left\{ \frac{\left( n-1\right) h_{k}^{2}}{1-h_{k}^{2}}>\frac{%
2\left( n-1\right) t^{2}/n}{1-2t^{2}/n}\right\} \leq \mathbb{P}\left\{ \frac{%
\left( n-1\right) h_{k}^{2}}{1-h_{k}^{2}}>\frac{2\left( n-1\right)
t^{2}/\left( n-2\right) }{1-t^{2}/\left( n-2\right) }\right\}  \\
&\leq &\mathbb{P}\left\{ \frac{\left( n-1\right) h_{k}^{2}}{1-h_{k}^{2}}%
>\left( n-1\right) \left( e^{2t^{2}/(n-2)}-1\right) \right\} \leq \left(
1+\epsilon _{n-1}\right) e^{-t^{2}}/\left( \pi ^{1/2}t\right) \text{,}
\end{eqnarray*}%
where the first inequality holds when $t^{2}\geq 2$, and the second
inequality follows from the fact $e^{x}-1\leq x/(1-\frac{x}{2})$ for $0<x<2$%
. Now let $t^{2}=\delta _{1}\log q>2$, and $\lambda _{1}=\left(
1+\varepsilon _{1}\right) \sqrt{\frac{2\delta _{1}\log q}{n}}$ with $\xi
=3/\varepsilon _{1}+1$, then we have $\lambda _{1}\frac{\xi -1}{\xi +1}%
\left( 1-\tau \right) >\sqrt{\frac{2\delta _{1}\log q}{n}}\ $and
\begin{eqnarray*}
\mathbb{P}\left\{ \cap _{i=1}^{4}I_{i}\right\}  &\geq &\mathbb{P}\left\{
\frac{\nu }{\theta ^{ora}}\leq \sqrt{\frac{2\delta _{1}\log q}{n}}\right\} -%
\mathbb{P}\left\{ \left( \cap _{i=2}^{4}I_{i}\right) ^{c}\right\}  \\
&\geq &1-q\cdot \mathbb{P}\left\{ \left\vert h_{k}\right\vert >\sqrt{\frac{%
2\delta _{1}\log q}{n}}\right\} -\mathbb{P}\left\{ \left( \cap
_{i=2}^{4}I_{i}\right) ^{c}\right\}  \\
&\geq &1-\left( \frac{1}{\sqrt{\pi \delta _{1}}}+o\left( 1\right) \right)
\frac{q^{-\delta _{1}+1}}{\sqrt{\log q}}\text{,}
\end{eqnarray*}%
which immediately implies $\mathbb{P}\left\{ I_{1}^{c}\right\} \leq O\left(
q^{-\delta _{1}+1}/\sqrt{\log q}\right) $.

\subsection{Proof of Theorem \protect\ref{theorem:step2}}

The whole proof is based on the results in Theorem \ref{theorem:step1}. In
particular with probability $1-o\left( p\cdot q^{-\delta _{1}+1}\right) $,
the following events hold,%
\begin{eqnarray}
\frac{1}{n}\left\Vert \mathbf{\hat{Z}}_{j}-\mathbf{Z}_{j}\right\Vert ^{2}
&\leq &C_{1}s_{1}\frac{\log q}{n}\text{ for all }j\text{,}
\label{proof2: pred error of Z} \\
\left\Vert \mathbf{\hat{\gamma}}_{j}-\mathbf{\gamma }_{j}\right\Vert _{1}
&\leq &C_{2}s_{1}\sqrt{\frac{\log q}{n}}\text{ for all }j\text{.}
\label{proof2: l1 error of gamma}
\end{eqnarray}%
From now on the analysis is conditioned on the two events above. This proof
is also based on the key lemma, Lemma \ref{KeyLemma:Scaled Lasso}. We apply
Lemma \ref{KeyLemma:Scaled Lasso} with $\mathbf{R=D}b^{true}+\mathbf{E}$
replaced by $\mathbf{\hat{Z}}_{m}\mathbf{=\hat{Z}}_{A^{c}}\beta _{m}+\mathbf{%
E}_{m}$ for each $m\in A=\left\{ i,j\right\} $, $\lambda $ replaced by $%
\lambda _{2}$ and sparsity $s$ replaced by $C_{\beta }s_{2},$ where $\mathbf{%
E}_{m}$ is defined by the regression model Equation (\ref{regression-data})
as%
\begin{eqnarray}
\mathbf{E}_{m} &\equiv &\mathbf{\epsilon }_{m}+\Delta _{Z_{m}}
\label{proof2: noisy noise E_m} \\
&=&\mathbf{\epsilon }_{m}+\left( \mathbf{\hat{Z}}_{m}-\mathbf{Z}_{m}\right)
+\left( \mathbf{Z}_{A^{c}}-\mathbf{\hat{Z}}_{A^{c}}\right) \beta _{m}\text{,}
\notag
\end{eqnarray}%
and $C_{\beta }=2M_{2}$ because the definition of $\beta _{m}$ in Equation (%
\ref{true beta}) implies it is a weighted sum of two columns of $\Omega $
with weight bounded by $M_{2}$.

To obtain our desired result $\left\vert \hat{\psi}_{kl}-\psi
_{kl}^{ora}\right\vert =\left\vert \mathbf{\epsilon }_{k}^{T}\mathbf{%
\epsilon }_{l}/n-\mathbf{\hat{\epsilon}}_{k}^{T}\mathbf{\hat{\epsilon}}%
_{l}/n\right\vert $ for each pair $k,l\in $ $A=\left\{ i,j\right\} $, it's
sufficient for us to bound $\left\vert \mathbf{\hat{\epsilon}}_{k}^{T}%
\mathbf{\hat{\epsilon}}_{l}/n-\mathbf{E}_{k}^{T}\mathbf{E}_{l}/n\right\vert $
and $\left\vert \mathbf{\epsilon }_{k}^{T}\mathbf{\epsilon }_{l}/n-\mathbf{E}%
_{k}^{T}\mathbf{E}_{l}/n\right\vert $ separately and then to apply the
triangle inequality. The following two lemmas are useful to establish those
two bounds.

\begin{lemma}
\label{lemma: step2_ prob 1} There exists some constant $C_{in}>0$ such that
$\left\vert \mathbf{\epsilon }_{k}^{T}\mathbf{\epsilon }_{l}/n-\mathbf{E}%
_{k}^{T}\mathbf{E}_{l}/n\right\vert \leq C_{in}\lambda _{1}^{2}s_{1}$ with
probability $1-o\left( q^{-\delta _{1}}\right) $.
\end{lemma}

\begin{lemma}
\label{lemma: step2_ prob 2} There exist some constants $C_{k}^{\prime }$, $%
1\leq k\leq 3$ such that for each $m\in A=\left\{ i,j\right\} $,
\begin{eqnarray*}
\left\Vert \beta _{m}-\hat{\beta}_{m}\right\Vert _{1} &\leq &C_{1}^{\prime
}\lambda _{2}s_{2}\text{,} \\
\left\Vert \mathbf{\hat{Z}}_{A^{c}}\left( \beta _{m}-\hat{\beta}_{m}\right)
\right\Vert ^{2}/n &\leq &C_{2}^{\prime }\lambda _{2}^{2}s_{2}\text{,} \\
\left\Vert \mathbf{\hat{Z}}_{A^{c}}^{T}\mathbf{E}_{m}/n\right\Vert _{\infty
} &\leq &C_{3}^{\prime }\lambda _{2}\text{,}
\end{eqnarray*}%
with probability $1-o\left( p^{-\delta _{2}+1}\right) $.
\end{lemma}

Before moving on, we point out a fact we will use several times in the proof,%
\begin{equation}
\left\Vert \beta _{m}\right\Vert _{1}\leq 2M_{2}^{2}\text{,}
\label{proof2: l1 bound of beta}
\end{equation}%
which follows from Equation (\ref{true beta}) and Condition $5$. Hence we
have
\begin{equation}
\left\Vert \Delta _{Z_{m}}\right\Vert /\sqrt{n}\leq \frac{\left\Vert \mathbf{%
\hat{Z}}_{m}-\mathbf{Z}_{m}\right\Vert }{\sqrt{n}}+\frac{\max_{k}\left\Vert
\mathbf{\hat{Z}}_{k}-\mathbf{Z}_{k}\right\Vert }{\sqrt{n}}\left\Vert \beta
_{m}\right\Vert _{1}\leq \sqrt{C_{3}s_{1}\lambda _{1}^{2}}\text{,}
\label{proof2: Delta Z bound}
\end{equation}%
for some constant $C_{3}>0$.

To bound the term $\left\vert \mathbf{\hat{\epsilon}}_{k}^{T}\mathbf{\hat{%
\epsilon}}_{l}/n-\mathbf{E}_{k}^{T}\mathbf{E}_{l}/n\right\vert $, we note
that for any $k,l\in $ $A=\left\{ i,j\right\} $,
\begin{eqnarray}
\left\vert \mathbf{\hat{\epsilon}}_{k}^{T}\mathbf{\hat{\epsilon}}_{l}/n-%
\mathbf{E}_{k}^{T}\mathbf{E}_{l}/n\right\vert  &=&\left\vert \left( \mathbf{E%
}_{k}+\mathbf{\hat{Z}}_{A^{c}}\left( \beta _{k}-\hat{\beta}_{k}\right)
\right) ^{T}\left( \mathbf{E}_{l}+\mathbf{\hat{Z}}_{A^{c}}\left( \beta _{l}-%
\hat{\beta}_{l}\right) \right) /n-\mathbf{E}_{k}^{T}\mathbf{E}%
_{l}/n\right\vert   \notag \\
&\leq &\left\Vert \mathbf{\hat{Z}}_{A^{c}}^{T}\mathbf{E}_{k}/n\right\Vert
_{\infty }\left\Vert \beta _{k}-\hat{\beta}_{k}\right\Vert _{1}+\left\Vert
\mathbf{\hat{Z}}_{A^{c}}^{T}\mathbf{E}_{l}/n\right\Vert _{\infty }\left\Vert
\beta _{l}-\hat{\beta}_{l}\right\Vert _{1}  \notag \\
&&+\left\Vert \mathbf{\hat{Z}}_{A^{c}}\left( \beta _{k}-\hat{\beta}%
_{k}\right) \right\Vert \cdot \left\Vert \mathbf{\hat{Z}}_{A^{c}}\left(
\beta _{l}-\hat{\beta}_{l}\right) \right\Vert /n  \notag \\
&\leq &\left( 2C_{1}^{\prime }C_{3}^{\prime }+C_{2}^{\prime }\right) \lambda
_{2}^{2}s_{2}\text{,}  \label{proof2: intermediate theta}
\end{eqnarray}%
where we applied Lemma \ref{lemma: step2_ prob 2} in the last inequality.

Lemma \ref{lemma: step2_ prob 1}, together with Equation (\ref{proof2:
intermediate theta}), immediately implies the desired result (\ref{result:
step2: close to oracle}),
\begin{equation*}
\left\vert \hat{\psi}_{kl}-\psi _{kl}^{ora}\right\vert \leq C_{4}^{\prime
}\left( s_{2}\frac{\log p}{n}+s_{1}\frac{\log q}{n}\right) \text{,}
\end{equation*}%
for some constant $C_{4}^{\prime }$ with probability $1-o\left( p\cdot
q^{-\delta _{1}+1}+p^{-\delta _{2}+1}\right) $. Since the spectrum of $\Psi
_{A,A}$ is bounded below by $M_{2}^{-1}$ and above by $M_{2}$ and the
functional $\left( \Omega _{A,A}\right) _{kl}=\left( \Psi _{A,A}^{-1}\right)
_{kl}$ is Lipschitz in a neighborhood of $\Psi _{A,A}$ for $k,l\in A$, we
obtain that Equation (\ref{result: step2: close to oracle omega}) is an
immediate consequence of Equation (\ref{result: step2: close to oracle})$.$
Note that $\omega _{ij}^{ora}$ is the MLE of $\omega _{ij}$ in the model $%
\left( \eta _{i},\eta _{j}\right) ^{T}\sim N\left( 0,\Omega
_{A,A}^{-1}\right) $ with three parameters given $n$ samples. Whenever $%
s_{2}=o\left( \frac{\sqrt{n}}{\log p}\right) $ and $s_{1}=o\left( \frac{%
\sqrt{n}}{\log q}\right) $, we have $s_{2}\frac{\log p}{n}+s_{1}\frac{\log q%
}{n}=o(\frac{1}{\sqrt{n}})$. Therefore we have $\omega _{ij}^{ora}-\hat{%
\omega}_{ij}=o_{p}\left( \omega _{ij}^{ora}-\omega _{ij}\right) $, which
immediately implies Equation (\ref{efficiency result}) in Theorem \ref%
{theorem:step2},
\begin{equation*}
\sqrt{nF_{ij}}\left( \hat{\omega}_{ij}-\omega _{ij}\right) \overset{D}{\sim }%
\sqrt{nF_{ij}}\left( \omega _{ij}^{ora}-\omega _{ij}\right) \overset{D}{%
\rightarrow }N\left( 0,1\right) \text{,}
\end{equation*}%
where $F_{ij}$ is the Fisher information of $\omega _{ij}$.

It is then enough to prove Lemma \ref{lemma: step2_ prob 1} and Lemma \ref%
{lemma: step2_ prob 2} to complete the proof of Theorem \ref{theorem:step2}.

\subsubsection{\textbf{Proof of Lemma \protect\ref{lemma: step2_ prob 1}}}

We show that $\left\vert \mathbf{\epsilon }_{k}^{T}\mathbf{\epsilon }_{l}/n-%
\mathbf{E}_{k}^{T}\mathbf{E}_{l}/n\right\vert \leq C_{in}\lambda
_{1}^{2}s_{1}$ with probability $1-o\left( q^{-\delta _{1}}\right) $ in this
section. By Equation (\ref{proof2: Delta Z bound}), we have
\begin{equation}
\left\vert \mathbf{\epsilon }_{k}^{T}\mathbf{\epsilon }_{l}/n-\mathbf{E}%
_{k}^{T}\mathbf{E}_{l}/n\right\vert \leq \left\vert \frac{\mathbf{\epsilon }%
_{k}^{T}\Delta _{Z_{l}}+\mathbf{\epsilon }_{l}^{T}\Delta _{Z_{k}}+\Delta
_{Z_{l}}^{T}\Delta _{Z_{k}}}{n}\right\vert \leq C_{3}s_{1}\lambda _{1}^{2}+%
\frac{\left\vert \mathbf{\epsilon }_{k}^{T}\Delta _{Z_{l}}\right\vert
+\left\vert \mathbf{\epsilon }_{l}^{T}\Delta _{Z_{k}}\right\vert }{n}\text{.}
\label{proof2:  intermediate theta,  easy part}
\end{equation}%
To bound the term $\left\vert \mathbf{\epsilon }_{k}^{T}\Delta
_{Z_{l}}\right\vert $, we note that by the definition of $\Delta _{Z_{l}}$
in Equation (\ref{proof2: noisy noise E_m}) there exists some constant $C_{4}
$ such that,%
\begin{eqnarray}
\frac{\left\vert \mathbf{\epsilon }_{k}^{T}\Delta _{Z_{l}}\right\vert }{n}
&=&\left\vert \mathbf{\epsilon }_{k}^{T}\left[ \left( \mathbf{\hat{Z}}_{l}-%
\mathbf{Z}_{l}\right) +\left( \mathbf{Z}_{A^{c}}-\mathbf{\hat{Z}}%
_{A^{c}}\right) \beta _{m}\right] \right\vert /n  \notag \\
&=&\left\vert \mathbf{\epsilon }_{k}^{T}\mathbf{X}\left[ \left( \gamma _{l}-%
\hat{\gamma}_{l}\right) +\left( \mathbf{\hat{\gamma}}_{A^{c}}-\mathbf{\gamma
}_{A^{c}}\right) \beta _{m}\right] \right\vert /n  \notag \\
&\leq &\left\Vert \mathbf{\epsilon }_{k}^{T}\mathbf{X}/n\right\Vert _{\infty
}\left\Vert \left( \gamma _{l}-\hat{\gamma}_{l}\right) +\left( \mathbf{\hat{%
\gamma}}_{A^{c}}-\mathbf{\gamma }_{A^{c}}\right) \beta _{m}\right\Vert _{1}
\notag \\
&\leq &\left\Vert \mathbf{\epsilon }_{k}^{T}\mathbf{X}/n\right\Vert _{\infty
}\max_{i}\left\Vert \gamma _{i}-\hat{\gamma}_{i}\right\Vert _{1}\left(
1+\left\Vert \beta _{m}\right\Vert _{1}\right) \leq \left\Vert \mathbf{%
\epsilon }_{k}^{T}\mathbf{X}/n\right\Vert _{\infty }C_{4}s_{1}\lambda _{1}%
\text{,}  \label{proof2: intermediate theta, easy part2}
\end{eqnarray}%
where the last inequality follows from Equations (\ref{proof2: l1 error of
gamma}) and (\ref{proof2: l1 bound of beta}). Since $\mathbf{\epsilon }_{k}%
\overset{i.i.d.}{\sim }$ $N\left( 0,\psi _{kk}\right) $ and $\mathbf{X}$ are
independent, it can be seen that each coordinate of $\mathbf{\epsilon }%
_{k}^{T}\mathbf{X}$ is a sum of $n$ i.i.d. sub-exponential variables with
bounded parameter under either Condition $3$ or $3^{\prime }$. A union bound
with $q$ coordinates and another application of Bernstein inequality in
Equation (\ref{Bernstein}) with $t=C_{5}\sqrt{\frac{\log q}{n}}$ imply that $%
\left\Vert \mathbf{\epsilon }_{k}^{T}\mathbf{X}/n\right\Vert _{\infty }\leq
C_{5}\sqrt{\frac{\log q}{n}}$ with probability $1-o\left( q^{-\delta
_{1}}\right) $ for some large constant $C_{5}>0$. This fact, together with
Equation (\ref{proof2: intermediate theta, easy part2}), implies that $%
\left\vert \mathbf{\epsilon }_{k}^{T}\Delta _{Z_{l}}\right\vert /n=O\left(
s_{1}\lambda _{1}^{2}\right) $ with probability $1-o\left( q^{-\delta
_{1}}\right) $. Similar result holds for $\left\vert \mathbf{\epsilon }%
_{l}^{T}\Delta _{Z_{k}}/n\right\vert $. Together with Equation (\ref{proof2:
intermediate theta, easy part}), this result completes our claim on $%
\left\vert \mathbf{\epsilon }_{k}^{T}\mathbf{\epsilon }_{l}/n-\mathbf{E}%
_{k}^{T}\mathbf{E}_{l}/n\right\vert $ and finishes the proof of Lemma \ref%
{lemma: step2_ prob 1}.

\subsubsection{\textbf{Proof of Lemma \protect\ref{lemma: step2_ prob 2}}}

The Lemma \ref{lemma: step2_ prob 2} is an immediate consequence of Lemma %
\ref{KeyLemma:Scaled Lasso} applied to $\mathbf{\hat{Z}}_{m}\mathbf{=\hat{Z}}%
_{A^{c}}\beta _{m}+\mathbf{E}_{m}$ for each $m\in A=\left\{ i,j\right\} $
with parameter $\lambda _{2}$ and sparsity $C_{\beta }s_{2}$. We check the
following conditions $I_{1}-I_{4}$ in the Lemma \ref{KeyLemma:Scaled Lasso}
hold with probability $1-o\left( p^{-\delta _{2}+1}\right) $ to finish our
proof. This part of the proof is similar to the proof of Lemma \ref{lemma:
step1_ prob}. We thus directly apply those facts already shown in the proof
of Lemma \ref{lemma: step1_ prob} whenever possible. Let

\begin{eqnarray*}
I_{1} &=&\left\{ \nu \leq \theta ^{ora}\lambda _{2}\frac{\xi -1}{\xi +1}%
\left( 1-\tau _{2}\right) \text{ for some }\xi >1\right\} \text{,} \\
I_{2} &=&\left\{ \frac{\left\Vert \mathbf{\hat{Z}}_{k}\right\Vert }{\sqrt{n}}%
\in \left[ 1/A_{1}^{\prime },A_{1}^{\prime }\right] \text{ for all }k\in
A^{c}\right\} \text{,} \\
I_{3} &=&\left\{ \theta ^{ora}\in \left[ 1/A_{2}^{\prime },A_{2}^{\prime }%
\right] \text{, where }\theta ^{ora}=\frac{\left\Vert \mathbf{E}%
_{m}\right\Vert }{\sqrt{n}}\text{.}\right\} \text{,} \\
I_{4} &=&\left\{
\begin{array}{c}
\frac{\mathbf{W}^{T}\mathbf{W}}{n}\text{ satisfies lower-RE with }\left(
\alpha _{1},\zeta \left( n,p\right) \right) \text{ s.t.} \\
s_{2}\zeta \left( n,p\right) 8\left( 1+\xi \right) ^{2}A_{1}^{\prime 2}\leq
\min \{\frac{\alpha _{1}}{2},1\}%
\end{array}%
\right\} \text{,}
\end{eqnarray*}%
where we can set $\xi =3/\varepsilon _{2}+1$ for the current setting, $%
A_{1}^{\prime }=$ $A_{2}^{\prime }=\sqrt{3M_{2}}$, $\alpha _{1}=\frac{1}{%
2M_{2}}$ and $\zeta \left( n,p\right) =o(1/s_{2})$. Let us still define $%
\mathbf{W=\hat{Z}}_{A^{c}}\cdot diag\left( \frac{\sqrt{n}}{\left\Vert
\mathbf{\hat{Z}}_{k}\right\Vert }\right) $ as the standardized $\mathbf{\hat{%
Z}}_{A^{c}}$ in the Lemma \ref{KeyLemma:Scaled Lasso} of Section \ref{se:
key lemma}. The strategy is to show that $\mathbb{P}\left\{
I_{1}^{c}\right\} \leq O\left( p^{-\delta _{2}+1}/\sqrt{\log p}\right) $ and
$\mathbb{P}\left\{ I_{i}^{c}\right\} \leq o(p^{-\delta _{2}})$ for $i=2$, $3$
and $4$, which completes our proof.

\textbf{(1).} To study $\mathbb{P}\left\{ I_{2}^{c}\right\} $ and $\mathbb{P}%
\left\{ I_{3}^{c}\right\} $, we note that $\frac{\left\Vert \mathbf{\hat{Z}}%
_{k}\right\Vert }{\sqrt{n}}\leq \frac{\left\Vert \mathbf{Z}_{k}\right\Vert }{%
\sqrt{n}}+\frac{\left\Vert \mathbf{\hat{Z}}_{k}-\mathbf{Z}_{k}\right\Vert }{%
\sqrt{n}}$, where $\mathbf{Z}_{k}\sim N\left( 0,\sigma _{kk}I\right) $ and $%
\frac{\left\Vert \mathbf{\hat{Z}}_{k}-\mathbf{Z}_{k}\right\Vert }{\sqrt{n}}%
=O\left( \sqrt{s_{1}\lambda _{1}^{2}}\right) =o(1)$ according to Equation (%
\ref{proof2: pred error of Z}). Similarly $\frac{\left\Vert \mathbf{E}%
_{m}\right\Vert }{\sqrt{n}}\leq \frac{\left\Vert \mathbf{\epsilon }%
_{m}\right\Vert }{\sqrt{n}}+\frac{\left\Vert \Delta _{Z_{m}}\right\Vert }{%
\sqrt{n}}$ where $\mathbf{\epsilon }_{m}\sim N\left( 0,\psi _{mm}I\right) $
and $\frac{\left\Vert \Delta _{Zm}\right\Vert }{\sqrt{n}}=o(1)$ from
Equation (\ref{proof2: Delta Z bound}). Noting that $\psi _{mm}$, $\sigma
_{kk}\in \left[ M_{2}^{-1},M_{2}\right] $, we use the same argument as that
for $\mathbb{P}\left\{ I_{3}^{c}\right\} $ in the proof of Lemma \ref{lemma:
step1_ prob} to obtain
\begin{eqnarray*}
\mathbb{P}\left\{ I_{2}^{c}\right\}  &\leq &\mathbb{P}\left\{ \frac{%
\left\Vert \mathbf{Z}_{k}\right\Vert }{\sqrt{n}}\notin \left[ 1/\sqrt{2M_{2}}%
,\sqrt{2M_{2}}\right] \text{ for some }k\right\} \leq o(p^{-\delta _{2}})%
\text{,} \\
\mathbb{P}\left\{ I_{3}^{c}\right\}  &\leq &\mathbb{P}\left\{ \frac{%
\left\Vert \mathbf{\epsilon }_{m}\right\Vert }{\sqrt{n}}\notin \left[ 1/%
\sqrt{2M_{2}},\sqrt{2M_{2}}\right] \right\} \leq o(p^{-\delta _{2}})\text{.}
\end{eqnarray*}

\textbf{(2)}. To study the lower-RE condition of $\frac{\mathbf{W}^{T}%
\mathbf{W}}{n}$, as what we did in the proof of Lemma \ref{lemma: step1_
prob} and Lemma \ref{lower RE Lemma}, we essentially need to study $\frac{%
\mathbf{\hat{Z}}_{A^{c}}^{T}\mathbf{\hat{Z}}_{A^{c}}}{n}$ and to show the
following fact%
\begin{equation*}
\left\vert v^{T}\left( \frac{\mathbf{\hat{Z}}_{A^{c}}^{T}\mathbf{\hat{Z}}%
_{A^{c}}}{n}-\Sigma _{A^{c},A^{c}}\right) v\right\vert \leq \frac{1}{LM_{2}}%
\text{ for all }v\in \mathbb{B}_{0}\left( 2C_{\beta }s_{2}\right) \text{
with }\left\Vert v\right\Vert =1\text{,}
\end{equation*}%
where $\lambda _{\min }\left( \Sigma _{A^{c},A^{c}}\right) \geq 1/M_{2}$.
Following the same line of the proof in Lemma \ref{lemma: step1_ prob} for
the lower-RE condition of $\frac{\mathbf{X}^{T}\mathbf{X}}{n}$ with
normality assumption on $\mathbf{Z}$ and sparsity assumption $s_{2}=o\left(
\sqrt{\frac{n}{\log p}}\right) $, we can obtain that with probability $%
1-o\left( p^{-\delta _{2}}\right) $,
\begin{equation}
\sup_{v\in \mathbb{B}_{0}\left( 2C_{\beta }s_{2}\right) }\left\vert
v^{T}\left( \frac{\mathbf{Z}_{A^{c}}^{T}\mathbf{Z}_{A^{c}}}{n}-\Sigma
_{A^{c},A^{c}}\right) v\right\vert \leq \frac{\left\Vert v\right\Vert ^{2}}{%
2LM_{2}}\text{.}  \label{proof2: known RE}
\end{equation}%
Therefore all we need to show in the current setting is that with
probability $1-$ $o\left( p^{-\delta _{2}}\right) $,
\begin{equation}
\left\vert v^{T}\left( \frac{\mathbf{\hat{Z}}_{A^{c}}^{T}\mathbf{\hat{Z}}%
_{A^{c}}}{n}-\frac{\mathbf{Z}_{A^{c}}^{T}\mathbf{Z}_{A^{c}}}{n}\right)
v\right\vert \leq \frac{1}{2LM_{2}}\text{ for all }v\in \mathbb{B}_{0}\left(
2C_{\beta }s_{2}\right) \text{ with }\left\Vert v\right\Vert =1\text{.}
\label{proof2: RE condition key part}
\end{equation}%
To show Equation (\ref{proof2: RE condition key part}), we notice%
\begin{equation*}
\left\vert v^{T}\left( \frac{\mathbf{\hat{Z}}_{A^{c}}^{T}\mathbf{\hat{Z}}%
_{A^{c}}}{n}-\frac{\mathbf{Z}_{A^{c}}^{T}\mathbf{Z}_{A^{c}}}{n}\right)
v\right\vert \leq 2\left\vert v^{T}\left( \frac{\mathbf{Z}_{A^{c}}^{T}\left(
\mathbf{\hat{Z}}_{A^{c}}-\mathbf{Z}_{A^{c}}\right) }{n}\right) v\right\vert +%
\frac{\left\Vert \left( \mathbf{\hat{Z}}_{A^{c}}-\mathbf{Z}_{A^{c}}\right)
v\right\Vert ^{2}}{n}\equiv D_{1}+D_{2}\text{.}
\end{equation*}

To control $D_{2}$, we find%
\begin{equation}
\sqrt{D_{2}}\leq \max_{k}\frac{\left\Vert \mathbf{\hat{Z}}_{k}-\mathbf{Z}%
_{k}\right\Vert }{\sqrt{n}}\left\Vert v\right\Vert _{1}\leq \sqrt{s_{1}}%
\lambda _{1}\sqrt{2C_{\beta }s_{2}}=o(1)\text{,}  \label{proof2: control D2}
\end{equation}%
by Equation (\ref{proof2: pred error of Z}), $\left\Vert v\right\Vert
_{1}\leq \sqrt{2C_{\beta }s_{2}}\left\Vert v\right\Vert $ and sparsity
assumptions (\ref{step2: sparsity assumptions}).

To control $D_{1}$, we find $D_{1}\leq \sqrt{D_{2}}\cdot \frac{\left\Vert
\mathbf{Z}_{A^{c}}v\right\Vert }{\sqrt{n}}$, which is $o(1)$ with
probability $1-$ $o\left( p^{-\delta _{2}}\right) $ by Equation (\ref%
{proof2: control D2}) and the following result $\frac{\left\Vert \mathbf{Z}%
_{A^{c}}v\right\Vert }{\sqrt{n}}=O(1)$. Equation (\ref{proof2: known RE})
implies that with probability $1-$ $o\left( p^{-\delta _{2}}\right) $,
\begin{equation*}
\frac{\left\Vert \mathbf{Z}_{A^{c}}v\right\Vert }{\sqrt{n}}\leq \frac{%
\left\Vert v\right\Vert }{\sqrt{2LM_{2}}}+\left\Vert \Sigma
_{A^{c},A^{c}}^{1/2}v\right\Vert =O(1)\text{ for all }\left\Vert
v\right\Vert =1\text{.}
\end{equation*}

\textbf{(3).} Finally we study the probability of event $I_{1}$. In the
current setting,
\begin{eqnarray*}
\nu  &=&\left\Vert \mathbf{W}^{T}\mathbf{E}_{m}/n\right\Vert _{\infty
}=\left\Vert \mathbf{W}^{T}\mathbf{\epsilon }_{m}/n\right\Vert _{\infty
}+O\left( \left\Vert \mathbf{W}^{T}\Delta _{Z_{m}}/n\right\Vert _{\infty
}\right) \text{,} \\
\theta ^{ora} &=&\frac{\left\Vert \mathbf{E}_{m}\right\Vert }{\sqrt{n}}=%
\frac{\left\Vert \mathbf{\epsilon }_{m}\right\Vert }{\sqrt{n}}+O\left( \frac{%
\left\Vert \Delta _{Z_{m}}\right\Vert }{\sqrt{n}}\right) \text{.}
\end{eqnarray*}%
Following the same line of the proof for event $I_{1}$ in Lemma \ref{lemma:
step1_ prob}, we obtain that with probability $1-o\left( p^{-\delta
_{2}+1}\right) $,
\begin{equation*}
\left\Vert \mathbf{W}^{T}\mathbf{\epsilon }_{m}/n\right\Vert _{\infty }<%
\frac{\left\Vert \mathbf{\epsilon }_{m}\right\Vert }{\sqrt{n}}\lambda _{2}%
\frac{\xi -1}{\xi +1}\left( 1-\tau _{2}\right) \text{. }
\end{equation*}%
Thus to prove event $I_{1}$=$\left\{ \nu \leq \theta ^{ora}\lambda _{2}\frac{%
\xi -1}{\xi +1}\left( 1-\tau _{2}\right) \right\} $ holds with desired
probability, we only need to show with probability $1-o\left( p^{-\delta
_{2}+1}\right) $,
\begin{equation}
\left\Vert \mathbf{W}^{T}\Delta _{Z_{m}}/n\right\Vert _{\infty }=o(\lambda
_{2})\text{ and }\frac{\left\Vert \Delta _{Z_{m}}\right\Vert }{\sqrt{n}}=o(1)%
\text{.}  \label{proof2: goal of I1}
\end{equation}

Equation (\ref{proof2: Delta Z bound}) immediately implies $\frac{\left\Vert
\Delta _{Zm}\right\Vert }{\sqrt{n}}=O\left( \sqrt{s_{1}\lambda _{1}^{2}}%
\right) =o(1)$ by the sparsity assumptions (\ref{step2: sparsity assumptions}%
). To study $\left\Vert \mathbf{W}^{T}\Delta _{Z_{m}}/n\right\Vert _{\infty }
$, we obtain that there exists some constant $C_{6}>0$ such that,
\begin{eqnarray}
\left\Vert \mathbf{W}^{T}\Delta _{Z_{m}}/n\right\Vert _{\infty }
&=&\left\Vert diag\left( \frac{\sqrt{n}}{\left\Vert \mathbf{\hat{Z}}%
_{k}\right\Vert }\right) \mathbf{\hat{Z}}_{A^{c}}^{T}\Delta
_{Z_{m}}/n\right\Vert _{\infty }  \notag \\
&\leq &A_{1}^{\prime }\left\Vert \mathbf{Z}_{A^{c}}^{T}+\left( \mathbf{\hat{Z%
}}_{A^{c}}^{T}-\mathbf{Z}_{A^{c}}^{T}\right) \Delta _{Z_{m}}/n\right\Vert
_{\infty }  \notag \\
&\leq &A_{1}^{\prime }\left( \left\Vert \mathbf{Z}_{A^{c}}^{T}\Delta
_{Z_{m}}/n\right\Vert _{\infty }+\max_{i}\left\Vert \left( \mathbf{\hat{Z}}%
_{i}-\mathbf{Z}_{i}\right) /\sqrt{n}\right\Vert \left\Vert \Delta _{Z_{m}}/%
\sqrt{n}\right\Vert \right)   \notag \\
&\leq &A_{1}^{\prime }\left\Vert \mathbf{Z}_{A^{c}}^{T}\Delta
_{Z_{m}}/n\right\Vert _{\infty }+C_{6}s_{1}\lambda _{1}^{2}\text{,}
\label{proof2: W delta z}
\end{eqnarray}%
where we used $\frac{\left\Vert \mathbf{\hat{Z}}_{k}\right\Vert }{\sqrt{n}}%
\in \left[ 1/A_{1}^{\prime },A_{1}^{\prime }\right] $ for all $k$ on $I_{2}$
in the first inequality and Equations (\ref{proof2: pred error of Z}) and (%
\ref{proof2: Delta Z bound}) in the last inequality. By the sparsity
assumptions (\ref{step2: sparsity assumptions}), we have $s_{1}\lambda
_{1}^{2}=o(\lambda _{2})$. Thus it's sufficient to show $\left\Vert \mathbf{Z%
}_{A^{c}}^{T}\Delta _{Z_{m}}/n\right\Vert _{\infty }=o(\lambda _{2})$ with
probability $1-o\left( p^{-\delta _{2}+1}\right) $. In fact,
\begin{eqnarray}
\left\Vert \mathbf{Z}_{A^{c}}^{T}\Delta _{Z_{m}}/n\right\Vert _{\infty }
&\leq &\left\Vert \mathbf{Z}_{A^{c}}^{T}\mathbf{X}\left[ \left( \mathbf{%
\gamma }_{m}-\mathbf{\hat{\gamma}}_{m}\right) +\left( \mathbf{\hat{\gamma}}%
_{A^{c}}-\mathbf{\gamma }_{A^{c}}\right) \beta _{m}\right] /n\right\Vert
_{\infty }  \notag \\
&\leq &\left\Vert \mathbf{Z}_{A^{c}}^{T}\mathbf{X}/n\right\Vert _{\infty
}\left\Vert \left[ \left( \mathbf{\gamma }_{m}-\mathbf{\hat{\gamma}}%
_{m}\right) +\left( \mathbf{\hat{\gamma}}_{A^{c}}-\mathbf{\gamma }%
_{A^{c}}\right) \beta _{m}\right] \right\Vert _{1}  \notag \\
&\leq &\left\Vert \mathbf{Z}_{A^{c}}^{T}\mathbf{X}/n\right\Vert _{\infty
}\max_{i}\left\Vert \mathbf{\hat{\gamma}}_{i}-\mathbf{\gamma }%
_{i}\right\Vert _{1}\left( 1+\left\Vert \beta _{m}\right\Vert _{1}\right)
\notag \\
&\leq &\left\Vert \mathbf{Z}_{A^{c}}^{T}\mathbf{X}/n\right\Vert _{\infty
}Cs_{1}\lambda _{1}=o\left( \left\Vert \mathbf{Z}_{A^{c}}^{T}\mathbf{X}%
/n\right\Vert _{\infty }\right) \text{,}  \label{proof2: Z deltaZ}
\end{eqnarray}%
where the last inequality follows from Equations (\ref{proof2: l1 error of
gamma}) and (\ref{proof2: l1 bound of beta}).

Since each $\mathbf{Z}_{k}^{T}\overset{i.i.d.}{\sim }$ $N\left( 0,\sigma
_{kk}\right) $ and is independent of $\mathbf{X}$, it can be seen that each
entry of $\mathbf{Z}_{A^{c}}^{T}\mathbf{X}$ is a sum of $n$ i.i.d.
sub-exponential variables with finite parameter under either Condition $3$
or $3^{\prime }$. A union bound with $pq$ entries and an application of
Bernstein inequality in Equation (\ref{Bernstein}) with $t=C_{7}\sqrt{\frac{%
\log \left( pq\right) }{n}}$ imply that $\left\Vert \mathbf{Z}_{A^{c}}^{T}%
\mathbf{X}/n\right\Vert _{\infty }\leq C_{7}\sqrt{\frac{\log \left(
pq\right) }{n}}$ with probability $1-o\left( \left( pq\right) ^{-\delta
_{1}}\right) $ for some large constant $C_{7}>0$. This result, together with
Equation (\ref{proof2: Z deltaZ}), implies that $\left\Vert \mathbf{Z}%
_{A^{c}}^{T}\Delta _{Z_{m}}/n\right\Vert _{\infty }=o(\lambda _{2})$ with
probability $1-o\left( p^{-\delta _{2}+1}\right) $. Now we finish the proof
of $\left\Vert \mathbf{W}^{T}\Delta _{Z_{m}}/n\right\Vert _{\infty
}=o(\lambda _{2})$ by Equation (\ref{proof2: W delta z}) and hence the proof
of the Equation (\ref{proof2: goal of I1}) with probability $1-o\left(
p^{-\delta _{2}+1}\right) $. This completes our proof of Lemma \ref{lemma:
step2_ prob 2}.

\section{A Key Lemma}

\label{se: key lemma}

The lemma of scaled lasso introduced in this section is deterministic in
nature. It could be applied in different settings in which the assumptions
of design matrix, response variables and noise distribution may vary, as
long as the conditions of the lemma are satisfied. Therefore it's a very
useful building block in the analysis of many different problems. In
particular, the main theorems of both steps are based on this key lemma. The
proof of this lemma is similar as that in \cite{SZH12-ScaledLaoo} and \cite%
{RenSunZhZh2013}, but we use the restrict eigenvalue condition for the gram
matrix instead of CIF condition to easily adapt to different settings of
design matrix for our probabilistic analysis.

Consider the following general scaled $l_{1}$ penalized regression problem.
Denote the $n$ by $p_{0}$ dimensional design matrix by $\mathbf{D}=\left(
\mathbf{D}_{1},\ldots ,\mathbf{D}_{p_{0}}\right) $\textbf{, }the $n$
dimensional response variable $\mathbf{R}=\left( R_{1},\ldots ,R_{n}\right)
^{T}$ and the noise variable $\mathbf{E}=\left( E_{1},\ldots ,E_{n}\right)
^{T}$. The scaled lasso estimator with tuning parameter $\lambda $ of the
regression%
\begin{equation}
\mathbf{R=D}b^{true}+\mathbf{E}\text{\textbf{,}}  \label{General Regression}
\end{equation}%
is defined as
\begin{equation}
\left\{ \hat{b},\hat{\theta}\right\} =\arg \min_{b\in \mathbb{R}%
^{p_{0}},\theta \in \mathbb{R}^{+}}\left\{ \frac{\left\Vert \mathbf{R}-%
\mathbf{D}b\right\Vert ^{2}}{2n\theta }+\frac{\theta }{2}+\lambda
\sum_{k=1}^{p_{0}}\frac{\left\Vert \mathbf{D}_{k}\right\Vert }{\sqrt{n}}%
\left\vert b_{k}\right\vert \right\} \text{,}
\label{General Scaled Lasso with RE}
\end{equation}%
where the sparsity $s$ of the true coefficient $b^{true}$ is defined as
follows,%
\begin{equation}
s=\Sigma _{j=1}^{p_{0}}\min \left\{ 1,\left\vert b_{j}^{true}\right\vert
/\lambda \right\} \text{,}  \label{General Scaled Lasso: Sparsity}
\end{equation}%
which is a generalization of exact sparseness (the number of nonzero
entries).

We first normalize each column of the design matrix $\mathbf{D}$ to make the
analysis cleaner by setting
\begin{equation}
d_{k}=\frac{\left\Vert \mathbf{D}_{k}\right\Vert }{\sqrt{n}}b_{k}\text{, \
and \ }\mathbf{W}=\mathbf{D}\cdot diag\left( \frac{\sqrt{n}}{\left\Vert
\mathbf{D}_{k}\right\Vert }\right) \text{,}  \label{normalization}
\end{equation}%
and then rewrite the model (\ref{General Regression}) and the penalized
procedure (\ref{General Scaled Lasso with RE}) as follows,%
\begin{equation}
\mathbf{R}=\mathbf{W}d^{true}+\mathbf{E}_{j}\text{,}
\label{normalized regression data}
\end{equation}%
and%
\begin{eqnarray}
L_{\lambda }\left( d,\theta \right)  &=&\frac{\left\Vert \mathbf{R}-\mathbf{W%
}d\right\Vert ^{2}}{2n\theta }+\frac{\theta }{2}+\lambda \left\Vert
d\right\Vert _{1}\text{,}  \label{objective function} \\
\left\{ \hat{d},\hat{\theta}\right\}  &=&\arg \min_{d\in \mathbb{R}%
^{p_{0}},\theta \in \mathbb{R}^{+}}L_{\lambda }\left( d,\theta \right) \text{%
,}  \label{standard Scaled Lasso}
\end{eqnarray}%
where the true coefficients and the estimator of the standardized scaled
lasso regression (\ref{objective function}) are $d_{k}^{true}=\frac{%
\left\Vert \mathbf{D}_{k}\right\Vert }{\sqrt{n}}b_{k}^{true}$ and $\hat{d}%
_{k}=\hat{b}_{k}\frac{\left\Vert \mathbf{D}_{k}\right\Vert }{\sqrt{n}}$
respectively$.$

For this standardized scaled lasso regression, we introduce some important
notation, including the lower-RE condition on the gram matrix $\frac{\mathbf{%
W}^{T}\mathbf{W}}{n}$. The oracle estimator $\theta ^{ora}$ of the noise
level can be defined as
\begin{equation}
\theta ^{ora}=\frac{\left\Vert \mathbf{R}-\mathbf{W}d^{true}\right\Vert }{%
\sqrt{n}}=\frac{\left\Vert \mathbf{E}\right\Vert }{\sqrt{n}}\text{.}
\label{sigma_oracle}
\end{equation}%
Let $\left\vert K\right\vert $ be the cardinality of an index set $K$.
Define $T$ as the index set of those large coefficients of $d^{true}$,
\begin{equation}
T=\left\{ k:\left\vert d_{k}^{true}\right\vert \geq \lambda \right\} \text{.
}  \label{T}
\end{equation}%
We say the gram matrix $\frac{\mathbf{W}^{T}\mathbf{W}}{n}$ satisfies a
\textbf{lower-RE condition} with curvature $\alpha _{1}>0$ and tolerance $%
\zeta \left( n,p_{0}\right) >0$ if%
\begin{equation}
\mu ^{T}\frac{\mathbf{W}^{T}\mathbf{W}}{n}\mu \geq \alpha _{1}\left\Vert \mu
\right\Vert ^{2}-\zeta \left( n,p_{0}\right) \left\Vert \mu \right\Vert
_{1}^{2}\text{ for all }\mu \in \mathbb{R}\text{.}  \label{Lower-RE}
\end{equation}%
Moreover, we define
\begin{eqnarray}
\nu  &=&\left\Vert \mathbf{W}^{T}\left( \mathbf{R}-\mathbf{W}d^{true}\right)
/n\right\Vert _{\infty }=\left\Vert \mathbf{W}^{T}\mathbf{E}/n\right\Vert
_{\infty }\text{,}  \label{v} \\
\tau  &=&s\lambda ^{2}\cdot A_{2}\frac{2\xi }{\left( 1+\xi \right) }%
C_{1}\left( \frac{3\theta ^{ora}}{2},4\xi A_{1},\alpha _{1}\right) \text{,}
\label{tau}
\end{eqnarray}%
with constants $\xi >1$, $A_{1}$ and $A_{2}$ introduced in Lemma \ref%
{KeyLemma:Scaled Lasso}. $C_{1}\left( \frac{3\theta ^{ora}}{2},4\xi
A_{1},\alpha _{1}\right) $ is a constant depending on $\frac{3\theta ^{ora}}{%
2}$, $\xi A_{1}$ and $\alpha _{1}$ with its definition in Equation (\ref%
{C1-appendix}). It is bounded above if $\frac{3\theta ^{ora}}{2}$ and $4\xi
A_{1}$ are bounded above and $\alpha _{1}$ is bounded below by some
universal constants, respectively.

With the notation we can state the key lemma as follows.

\begin{lemma}
\label{KeyLemma:Scaled Lasso}Consider the scaled $l_{1}$ penalized
regression procedure (\ref{General Scaled Lasso with RE}). Whenever there
exist constants $\xi >1$, $A_{1}$ and $A_{2}$ such that the following
conditions are satisfied

\begin{enumerate}
\item $\nu \leq \theta ^{ora}\lambda \frac{\xi -1}{\xi +1}\left( 1-\tau
\right) $ for some $\xi >1$ and $\tau \leq 1/2$ defined in (\ref{tau});

\item $\frac{\left\Vert \mathbf{D}_{k}\right\Vert }{\sqrt{n}}\in \left[
1/A_{1},A_{1}\right] $ for all $k$;

\item $\theta ^{ora}\in \left[ 1/A_{2},A_{2}\right] $;

\item $\frac{\mathbf{W}^{T}\mathbf{W}}{n}$satisfies the lower-RE condition%
\textbf{\ }with $\alpha _{1}$ and $\zeta \left( n,p_{0}\right) $ such that%
\begin{equation}
s\zeta \left( n,p_{0}\right) 8\left( 1+\xi \right) ^{2}A_{1}^{2}\leq \min \{%
\frac{\alpha _{1}}{2},1\}\text{,}  \label{Good Lower-RE condition}
\end{equation}%
we have the following deterministic bounds
\begin{eqnarray}
\left\vert \hat{\theta}-\theta ^{ora}\right\vert  &\leq &C_{1}\lambda ^{2}s%
\text{,}  \label{Lemma result1} \\
\left\Vert \hat{b}-b^{true}\right\Vert ^{2} &\leq &C_{2}\lambda ^{2}s\text{,}
\label{Lemma result2} \\
\left\Vert \hat{b}-b^{true}\right\Vert _{1} &\leq &C_{3}\lambda s\text{,}
\label{Lemma result3} \\
\left\Vert \mathbf{D}\left( \hat{b}-b^{true}\right) \right\Vert ^{2}/n &\leq
&C_{4}\lambda ^{2}s\text{,}  \label{Lemma result4} \\
\left\Vert \mathbf{D}^{T}\mathbf{E}/n\right\Vert _{\infty } &\leq
&C_{5}\lambda \text{,}  \label{Lemma result5}
\end{eqnarray}%
where constants $C_{i}$ ($i=1,\ldots ,5$) only depend on $A_{1}$, $A_{2}$, $%
\alpha _{1}$ and $\xi $.
\end{enumerate}
\end{lemma}

\section{Appendix}

\label{se: appdix}

\subsection{Proof of Lemma \protect\ref{KeyLemma:Scaled Lasso}}

The function $L_{\lambda }\left( d,\theta \right) $ in Equation (\ref%
{objective function}) is jointly convex in $\left( d,\theta \right) $. For
fixed $\theta >0$, denote the minimizer of $L_{\lambda }\left( d,\theta
\right) $ over all $d\in \mathbb{R}^{p_{0}}$ by $\hat{d}\left( \theta
\lambda \right) $, a function of $\theta \lambda $, i.e.,
\begin{equation}
\hat{d}\left( \theta \lambda \right) =\arg \min_{d\in \mathbb{R}%
^{p_{0}}}L_{\lambda }\left( d,\theta \right) =\arg \min_{d\in \mathbb{R}%
^{p_{0}}}\left\{ \frac{\left\Vert \mathbf{R}-\mathbf{W}d\right\Vert ^{2}}{2n}%
+\lambda \theta \left\Vert d\right\Vert _{1}\right\} \text{,}
\label{Ordinary Lasso}
\end{equation}%
then if we knew $\hat{\theta}$ in the solution of Equation (\ref{standard
Scaled Lasso}), the solution for the equation is $\left\{ \hat{d}\left( \hat{%
\theta}\lambda \right) ,\hat{\theta}\right\} $. We recognize that $\hat{d}%
\left( \hat{\theta}\lambda \right) $ is just the standard lasso with the
penalty $\hat{\theta}\lambda ,$ however we don't know the estimator $\hat{%
\theta}$. The strategy of our analysis is that we first show that $\hat{%
\theta}$ is very close to its oracle estimator $\theta ^{ora}$, then the
standard lasso analysis would imply the desired result Equations (\ref{Lemma
result2})-(\ref{Lemma result5}) under the assumption that $\hat{\theta}%
/\theta ^{ora}=1+O(\lambda ^{2}s)$. For the standard lasso analysis, some
kind of regularity condition is assumed on the design matrix $\mathbf{W}^{T}%
\mathbf{W/}n$ in the regression literature. In this paper we use the
lower-RE condition, which is one of the most general conditions.

Let $\mu =\lambda \theta $. From the Karush-Kuhn-Tucker condition, $\hat{d}%
\left( \mu \right) $ is the solution to the Equation (\ref{Ordinary Lasso})
if and only if%
\begin{eqnarray}
\mathbf{W}_{k}^{T}\left( \mathbf{R}-\mathbf{W}\hat{d}\left( \mu \right)
\right) /n &=&\mu \cdot sgn\left( \hat{d}_{k}\left( \mu \right) \right)
\text{,}\mbox{\rm  \ if }\hat{d}_{k}\left( \mu \right) \neq 0\text{,}
\label{KKT} \\
\mathbf{W}_{k}^{T}\left( \mathbf{R}-\mathbf{W}\hat{d}\left( \mu \right)
\right) /n &\in &\left[ -\mu ,\mu \right] \text{,}\mbox{\rm  \ if }\hat{d}%
_{k}\left( \mu \right) =0\text{.}  \notag
\end{eqnarray}

Let $C_{2}\left( a_{1},a_{2}\right) $ and $C_{1}\left(
a_{11},a_{12},a_{2}\right) $ be constants depending on $a_{1}$,
$a_{2}$ and $a_{11}$, $a_{12}$, $a_{2}$, respectively. The constant $C_{2}$ is bounded above if $a_{1}$ is bounded above and $a_{2}$ is
bounded below by constants, respectively. The constant $C_{1}$ is bounded above whenever $a_{11}$ and $a_{12}$ are bounded
above and $a_{2}$ is bounded below by constants. The explicit
formulas of $C_{1}$ and $C_{2}$ are given as follows,
\begin{eqnarray}
C_{2}\left( a_{1},a_{2}\right)  &=&\frac{a_{1}}{a_{2}}+\left[ \left( \frac{%
a_{1}}{a_{2}}\right) ^{2}+\frac{2\left( a_{1}+1\right) }{a_{2}}\right] ^{1/2}%
\text{,}  \notag \\
C_{1}\left( a_{11},a_{12},a_{2}\right)  &=&a_{12}\left( 1+C_{2}\left(
a_{11}\times a_{12},a_{2}\right) \right) \text{.}  \label{C1-appendix}
\end{eqnarray}%
The following propositions are helpful to establish our result. The proof is
given in Sections \ref{se: proof of Prop1} and \ref{se: proof of Prop 2}.

\begin{proposition}
\label{Standard Lasso Result} The sparsity $s$ is defined in Equation (\ref%
{General Scaled Lasso: Sparsity}). For any $\xi >1$, assuming $\nu \leq \mu
\frac{\xi -1}{\xi +1}$ and conditions $2$, $4$ in Lemma \ref{KeyLemma:Scaled
Lasso} hold, we have%
\begin{eqnarray}
\left\Vert \hat{d}\left( \mu \right) -d^{true}\right\Vert _{1} &\leq
&C_{1}\left( \frac{\mu }{\lambda },4\xi A_{1},\alpha _{1}\right) \lambda s%
\text{,}  \label{standard Lasso Result: L1 norm bound} \\
\left\Vert \hat{d}\left( \mu \right) -d^{true}\right\Vert  &\leq
&C_{2}\left( 4\xi A_{1}\frac{\mu }{\lambda },\alpha _{1}\right) \lambda
\sqrt{s}\text{,}  \label{standard Lasso Result: L2 norm bound} \\
\frac{1}{n}\left\Vert \mathbf{W}\left( d^{true}-\hat{d}\left( \mu \right)
\right) \right\Vert ^{2} &\leq &\left( \nu +\mu \right) \left\Vert \hat{d}%
\left( \mu \right) -d^{true}\right\Vert _{1}\text{.}
\label{standard Lasso Result: prediction bound}
\end{eqnarray}
\end{proposition}

\begin{proposition}
\label{Error bound of sigma} Let $\left\{ \hat{d},\hat{\theta}\right\} $ be
the solution of the scaled lasso (\ref{standard Scaled Lasso}). For any $\xi
>1$, assuming conditions $1-4$ in Lemma \ref{KeyLemma:Scaled Lasso} hold$,$
then we have%
\begin{equation}
\left\vert \frac{\hat{\theta}}{\theta ^{ora}}-1\right\vert \leq \tau \text{.}
\label{Error bound}
\end{equation}
\end{proposition}

Now we finish our proof with these two propositions. According to Conditions
$1-4$ in Lemma \ref{KeyLemma:Scaled Lasso} , Proposition \ref{Error bound of
sigma} implies $\nu \leq \mu \frac{\xi -1}{\xi +1}$ with $\mu =\lambda \hat{%
\theta}$ and Proposition \ref{Standard Lasso Result} further implies that
there exist some constants $c_{1},$ $c_{2}$ and $c_{3}$ such that \
\begin{eqnarray*}
\left\Vert \hat{d}\left( \lambda \hat{\theta}\right) -d^{true}\right\Vert
_{1} &\leq &c_{1}\lambda s\text{,} \\
\left\Vert \hat{d}\left( \lambda \hat{\theta}\right) -d^{true}\right\Vert
&\leq &c_{2}\lambda \sqrt{s}\text{,} \\
\frac{\left\Vert \mathbf{W}\left( d^{true}-\hat{d}\left( \lambda \hat{\theta}%
\right) \right) \right\Vert ^{2}}{n} &\leq &c_{3}\lambda ^{2}s\text{.}
\end{eqnarray*}%
Note that $\frac{\mu }{\lambda }=\hat{\theta}\in \left[ \theta ^{ora}(1-\tau
),\theta ^{ora}(1+\tau )\right] \subset \left[ \frac{1}{2A_{2}},\frac{3A_{2}%
}{2}\right] $. Thus the constants $c_{1}$, $c_{2}$ and $c_{3}$ only depends
on $A_{1}$, $A_{2}$, $\alpha _{1}$ and $\xi $. Now we transfer the results
above on standardized scaled lasso (\ref{standard Scaled Lasso}) back to the
general scaled lasso (\ref{General Scaled Lasso with RE}) through the
bounded scaling constants $\left\{ \frac{\sqrt{n}}{\left\Vert \mathbf{D}%
_{k}\right\Vert }\right\} $ and immediately have the desired results (\ref%
{Lemma result2})-(\ref{Lemma result4}). Result (\ref{Lemma result1}) is an
immediate consequence of Proposition \ref{Error bound of sigma} and Result (%
\ref{Lemma result5}) is an immediate consequence of assumptions $1-3$.

\subsection{Proof of Proposition \protect\ref{Standard Lasso Result}}

\label{se: proof of Prop1}

Notice that%
\begin{eqnarray}
&&\frac{1}{n}\left\Vert \mathbf{W}\left( d^{true}-\hat{d}\left( \mu \right)
\right) \right\Vert ^{2}  \notag \\
&=&\frac{\left( d^{true}-\hat{d}\left( \mu \right) \right) ^{T}\left(
\mathbf{W}^{T}\left( \mathbf{R}-\mathbf{W}\hat{d}\left( \mu \right) \right) -%
\mathbf{W}^{T}\left( \mathbf{R}-\mathbf{W}d^{true}\right) \right) }{n}
\notag \\
&\leq &\mu \left( \left\Vert d^{true}\right\Vert _{1}-\left\Vert \hat{d}%
\left( \mu \right) \right\Vert _{1}\right) +\nu \left\Vert d^{true}-\hat{d}%
\left( \mu \right) \right\Vert _{1}  \label{appendix 3} \\
&\leq &\left( \mu +\nu \right) \left\Vert \left( d^{true}-\hat{d}\left( \mu
\right) \right) _{T}\right\Vert _{1}+2\mu \left\Vert \left( d^{true}\right)
_{T^{c}}\right\Vert _{1}-\left( \mu -\nu \right) \left\Vert \left( d^{true}-%
\hat{d}\left( \mu \right) \right) _{T^{c}}\right\Vert _{1}\text{,}
\label{appendix4}
\end{eqnarray}%
where the first inequality follows from the KKT conditions (\ref{KKT}).

Now define $\Delta =\hat{d}\left( \mu \right) -d^{true}.$ Equation (\ref%
{appendix 3}) also implies the desired inequality (\ref{standard Lasso
Result: prediction bound})
\begin{equation}
\Delta ^{T}\frac{\mathbf{W}^{T}\mathbf{W}}{n}\Delta \leq \left( \mu +\nu
\right) \left\Vert \Delta \right\Vert _{1}  \label{appendix 5}
\end{equation}%
We will first show that%
\begin{equation}
\left\Vert \Delta _{T^{c}}\right\Vert _{1}\leq \max \left\{ 2\left( 1+\xi
\right) \left\Vert \left( d^{true}\right) _{T^{c}}\right\Vert _{1},2\xi
\left\Vert \Delta _{T}\right\Vert _{1}\right\} \text{,}
\label{Cone condition}
\end{equation}%
then we are able to apply the lower-RE condition (\ref{Lower-RE}) to derive
the desired results.

To show Equation (\ref{Cone condition}), we note that our assumption $\nu
\leq \mu \frac{\xi -1}{\xi +1}$ with Equation (\ref{appendix4}) implies
\begin{equation}
\frac{1}{n}\left\Vert \mathbf{W}\Delta \right\Vert ^{2}\leq 2\mu \left(
\frac{\xi \left\Vert \Delta _{T}\right\Vert _{1}}{\xi +1}+\left\Vert \left(
d^{true}\right) _{T^{c}}\right\Vert _{1}-\frac{\left\Vert \Delta
_{T^{c}}\right\Vert _{1}}{\xi +1}\right) \text{.}
\label{appendix intermediate 1}
\end{equation}%
Suppose that
\begin{equation}
\left\Vert \Delta _{T^{c}}\right\Vert _{1}\geq 2\left( 1+\xi \right)
\left\Vert \left( d^{true}\right) _{T^{c}}\right\Vert _{1}\text{,}
\label{appendix 2}
\end{equation}%
then the inequality (\ref{appendix intermediate 1}) becomes%
\begin{equation*}
0\leq \frac{1}{n}\left\Vert \mathbf{W}\Delta \right\Vert ^{2}\leq \frac{\mu
}{\xi +1}\left( 2\xi \left\Vert \Delta _{T}\right\Vert _{1}-\left\Vert
\Delta _{T^{c}}\right\Vert _{1}\right) \text{,}
\end{equation*}%
which implies
\begin{equation}
\left\Vert \Delta _{T^{c}}\right\Vert _{1}\leq 2\xi \left\Vert \Delta
_{T}\right\Vert _{1}\text{.}  \label{appendix 1}
\end{equation}%
Therefore the complement of inequality (\ref{appendix 2}) and Equation (\ref%
{appendix 1}) together finish our proof of Equation (\ref{Cone condition}).

Before proceeding, we point out two facts which will be used below several
times. Note the sparseness $s$ is defined in terms of the true coefficients $%
b^{true}$ in Equation (\ref{General Scaled Lasso: Sparsity}) before
standardization but the index set $T$ is defined in term of $d^{true}$ in
Equation (\ref{T}) after standardization. Condition $2$ implies that this
standardization step doesn't change the sparseness up to a factor $A_{1}$.
Hence it's not hard to see that $\left\vert T\right\vert \leq A_{1}s$ and $%
\left\Vert \left( d^{true}\right) _{T^{c}}\right\Vert _{1}\leq A_{1}\lambda
s.$

Now we are able to apply the lower-RE condition of $\frac{\mathbf{W}^{T}%
\mathbf{W}}{n}$ to Equation (\ref{appendix 5}) and obtain that%
\begin{eqnarray*}
\Delta ^{T}\frac{\mathbf{W}^{T}\mathbf{W}}{n}\Delta  &\geq &\alpha
_{1}\left\Vert \Delta \right\Vert ^{2}-\zeta \left( n,p_{0}\right)
\left\Vert \Delta \right\Vert _{1}^{2} \\
&\geq &\alpha _{1}\left\Vert \Delta \right\Vert ^{2}-\zeta \left(
n,p_{0}\right) 8\left( 1+\xi \right) ^{2}\left( \left\Vert \left(
d^{true}\right) _{T^{c}}\right\Vert _{1}^{2}+\left\Vert \Delta
_{T}\right\Vert _{1}^{2}\right)  \\
&\geq &\left( \alpha _{1}-\left\vert T\right\vert \zeta \left(
n,p_{0}\right) 8\left( 1+\xi \right) ^{2}\right) \left\Vert \Delta
\right\Vert ^{2}-\zeta \left( n,p_{0}\right) 8\left( 1+\xi \right)
^{2}A_{1}^{2}\lambda ^{2}s^{2} \\
&\geq &\frac{\alpha _{1}}{2}\left\Vert \Delta \right\Vert ^{2}-\lambda ^{2}s%
\text{,}
\end{eqnarray*}%
where in the second, third and last inequalities we applied the facts (\ref%
{Cone condition}), $\left\Vert \Delta _{T}\right\Vert _{1}^{2}\leq
\left\vert T\right\vert \left\Vert \Delta _{T}\right\Vert ^{2}$, $\left\vert
T\right\vert \leq A_{1}s$, $\left\Vert \left( d^{true}\right)
_{T^{c}}\right\Vert _{1}\leq A_{1}\lambda s$ and $s\zeta \left(
n,p_{0}\right) 8\left( 1+\xi \right) ^{2}A_{1}^{2}\leq \min \{\frac{\alpha
_{1}}{2},1\}$. Moreover, by applying those facts used in last equation
again, we have%
\begin{eqnarray*}
\left( \mu +\nu \right) \left\Vert \Delta \right\Vert _{1} &\leq &4\xi \mu
\left( \left\Vert \left( d^{true}\right) _{T^{c}}\right\Vert _{1}+\left\Vert
\Delta _{T}\right\Vert _{1}\right)  \\
&\leq &4\xi \mu \left( A_{1}\lambda s+\sqrt{\left\vert T\right\vert }%
\left\Vert \Delta \right\Vert \right)  \\
&\leq &4\xi A_{1}\frac{\mu }{\lambda }\left( \lambda ^{2}s+\sqrt{s}\lambda
\left\Vert \Delta \right\Vert \right) \text{.}
\end{eqnarray*}%
The above two inequalities together with Equation (\ref{appendix 5}) imply
that%
\begin{equation*}
4\xi A_{1}\frac{\mu }{\lambda }\left( \lambda ^{2}s+\sqrt{s}\lambda
\left\Vert \Delta \right\Vert \right) \geq \frac{\alpha _{1}}{2}\left\Vert
\Delta \right\Vert ^{2}-\lambda ^{2}s\text{.}
\end{equation*}%
Define $S_{u}=4\xi A_{1}\frac{\mu }{\lambda }$. Some algebra about this
quadratic inequality implies the bound of $\Delta $ under $l_{2}$ norm
\begin{eqnarray*}
\left\Vert \Delta \right\Vert  &\leq &\left( \frac{S_{u}}{\alpha _{1}}+\left[
\left( \frac{S_{u}}{\alpha _{1}}\right) ^{2}+\frac{2\left( S_{u}+1\right) }{%
\alpha _{1}}\right] ^{1/2}\right) \lambda \sqrt{s} \\
&\equiv &C_{2}\left( 4\xi A_{1}\frac{\mu }{\lambda },\alpha _{1}\right)
\lambda \sqrt{s}\text{.}
\end{eqnarray*}

Combining this fact with Equation (\ref{Cone condition}), we finally obtain
the bound under $l_{1}$ norm (\ref{standard Lasso Result: L1 norm bound})%
\begin{eqnarray*}
\left\Vert \Delta \right\Vert _{1} &\leq &\left\Vert \Delta _{T}\right\Vert
_{1}+\left\Vert \Delta _{T^{c}}\right\Vert _{1}\leq 2\left( 1+\xi \right)
\left( \left\Vert \left( d^{true}\right) _{T^{c}}\right\Vert _{1}+\left\Vert
\Delta _{T}\right\Vert _{1}\right)  \\
&\leq &2\left( 1+\xi \right) \left( A_{1}s\lambda +\sqrt{\left\vert
T\right\vert }\left\Vert \Delta \right\Vert \right)  \\
&\leq &4\xi A_{1}\left( 1+C_{2}\left( 4\xi A_{1}\frac{\mu }{\lambda },\alpha
_{1}\right) \right) s\lambda  \\
&\equiv &C_{1}\left( \frac{\mu }{\lambda },4\xi A_{1},\alpha _{1}\right)
s\lambda \text{.}
\end{eqnarray*}

\subsection{Proof of Proposition \protect\ref{Error bound of sigma}}

\label{se: proof of Prop 2}

For $\tau $ defined in Equation (\ref{tau}), we need to show that $\hat{%
\theta}\geq \theta ^{ora}\left( 1-\tau \right) $ and $\hat{\theta}\leq
\theta ^{ora}\left( 1+\tau \right) $ on the event $\left\{ \nu \leq \theta
^{ora}\lambda \frac{\xi -1}{\xi +1}\left( 1-\tau \right) \right\} $. Let $%
\hat{d}\left( \theta \lambda \right) $ be the solution of (\ref{Ordinary
Lasso}) as a function of $\theta $, then
\begin{equation}
\frac{\partial }{\partial \theta }L_{\lambda }\left( \hat{d}\left( \theta
\lambda \right) ,\theta \right) =\frac{1}{2}-\frac{\left\Vert \mathbf{R}-%
\mathbf{W}\hat{d}\left( \theta \lambda \right) \right\Vert ^{2}}{2n\theta
^{2}}\text{,}  \label{Derivative}
\end{equation}%
since $\left\{ \frac{\partial }{\partial d}L_{\lambda }\left( d,\theta
\right) |_{d=\hat{d}\left( \theta \lambda \right) }\right\} _{k}=0$ for all $%
\hat{d}_{k}\left( \theta \lambda \right) \neq 0$, and $\left\{ \frac{%
\partial }{\partial \theta }\hat{d}\left( \theta \lambda \right) \right\}
_{k}=0$ for all $\hat{d}_{k}\left( \theta \lambda \right) =0$ which follows
from the fact that $\left\{ k:\hat{d}_{k}\left( \theta \lambda \right)
=0\right\} $ is unchanged in a neighborhood of $\theta $ for almost all $%
\theta $. Equation (\ref{Derivative}) plays a key in the proof.

\textbf{(1)}. \ To show that $\hat{\theta}\geq \theta ^{ora}\left( 1-\tau
\right) $ it's enough to show
\begin{equation*}
\frac{\partial }{\partial \theta }L_{\lambda }\left( \hat{d}\left( \theta
\lambda \right) ,\theta \right) |_{\theta =t_{1}}<0\text{,}
\end{equation*}%
where $t_{1}=\theta ^{ora}\left( 1-\tau \right) $, due to the strict
convexity of the objective function $L_{\lambda }\left( d,\theta \right) $
in $\theta $. Equation (\ref{Derivative}) implies that
\begin{eqnarray}
2t_{1}^{2}\frac{\partial }{\partial \theta }L_{\lambda }\left( \hat{d}\left(
\theta \lambda \right) ,\theta \right) |_{\theta =t_{1}} &=&t_{1}^{2}-\frac{%
\left\Vert \mathbf{R}-\mathbf{W}\hat{d}\left( t_{1}\lambda \right)
\right\Vert ^{2}}{n}  \notag \\
&\leq &t_{1}^{2}-\frac{\left\Vert \mathbf{R}-\mathbf{W}d^{true}+\mathbf{W}%
\left( \hat{d}\left( t_{1}\lambda \right) -d^{true}\right) \right\Vert ^{2}}{%
n}  \notag \\
&\leq &t_{1}^{2}-\left( \theta ^{ora}\right) ^{2}+2\left( d^{true}-\hat{d}%
\left( t_{1}\lambda \right) \right) ^{T}\frac{\mathbf{W}^{T}\left( \mathbf{R}%
-\mathbf{W}d^{true}\right) }{n}  \notag \\
&\leq &2t_{1}\left( t_{1}-\theta ^{ora}\right) +2\nu \left\Vert d^{true}-%
\hat{d}\left( t_{1}\lambda \right) \right\Vert _{1}\text{.}  \label{part1}
\end{eqnarray}%
On the event $\left\{ \nu \leq t_{1}\lambda \frac{\xi -1}{\xi +1}\right\}
=\left\{ \nu /\theta ^{ora}<\lambda \frac{\xi -1}{\xi +1}\left( 1-\tau
\right) \right\} $ we have%
\begin{eqnarray*}
&&2t_{1}^{2}\frac{\partial }{\partial \theta }L_{\lambda }\left( \hat{d}%
\left( \theta \lambda \right) ,\theta \right) |_{\theta =t_{1}} \\
&\leq &2t_{1}\left( t_{1}-\theta ^{ora}\right) +2t_{1}\lambda \frac{\xi -1}{%
\xi +1}\left\Vert d^{true}-\hat{d}\left( t_{1}\lambda \right) \right\Vert
_{1} \\
&\leq &2t_{1}\left[ -\tau \theta ^{ora}+\lambda \frac{\xi -1}{\xi +1}%
\left\Vert d^{true}-\hat{d}\left( t_{1}\lambda \right) \right\Vert _{1}%
\right] <0\text{.}
\end{eqnarray*}%
The last inequality follows from the definition of $\tau $ and the $l_{1}$
error bound in Equation (\ref{standard Lasso Result: L1 norm bound}) of
Proposition \ref{Standard Lasso Result}. Note that for $\lambda ^{2}s$
sufficiently small, we have small $\tau <1/2$. In fact, although $\left\Vert
d^{true}-\hat{d}\left( t_{1}\lambda \right) \right\Vert _{1}$ also depends
on $\tau $, our choice of $\tau $ is well-defined and is larger than $\frac{%
\lambda }{\theta ^{ora}}\frac{3(\xi -1)}{2(\xi +1)}\left\Vert d^{true}-\hat{d%
}\left( t_{1}\lambda \right) \right\Vert _{1}$.

\textbf{(2)}. Let $t_{2}=\theta ^{ora}\left( 1+\tau \right) $. To show the
other side $\hat{\theta}\leq \theta ^{ora}\left( 1+\tau \right) $ it is
enough to show%
\begin{equation*}
\frac{\partial }{\partial \theta }L_{\lambda }\left( \hat{d}\left( \theta
\lambda \right) ,\theta \right) |_{\theta =t_{2}}>0\text{.}
\end{equation*}%
Equation (\ref{Derivative}) implies that on the event $\left\{ \nu \leq
t_{2}\lambda \frac{\xi -1}{\xi +1}\right\} =\left\{ \nu /\theta
^{ora}<\lambda \frac{\xi -1}{\xi +1}\left( 1+\tau \right) \ \right\} $ we
have%
\begin{eqnarray*}
2t_{2}^{2}\frac{\partial }{\partial \theta }L_{\lambda }\left( \hat{d}\left(
\theta \lambda \right) ,\theta \right) |_{\theta =t_{2}} &=&t_{2}^{2}-\frac{%
\left\Vert \mathbf{R}-\mathbf{W}\hat{d}\left( t_{2}\lambda \right)
\right\Vert ^{2}}{n} \\
&=&t_{2}^{2}-\left( \theta ^{ora}\right) ^{2}+\left( \theta ^{ora}\right)
^{2}-\frac{\left\Vert \mathbf{R}-\mathbf{W}\hat{d}\left( t_{2}\lambda
\right) \right\Vert ^{2}}{n} \\
&=&t_{2}^{2}-\left( \theta ^{ora}\right) ^{2}+\frac{\left\Vert \mathbf{R}-%
\mathbf{W}d^{true}\right\Vert ^{2}-\left\Vert \mathbf{R}-\mathbf{W}\hat{d}%
\left( t_{2}\lambda \right) \right\Vert ^{2}}{n} \\
&=&t_{2}^{2}-\left( \theta ^{ora}\right) ^{2}+\frac{\left( \hat{d}\left(
t_{2}\lambda \right) -d^{true}\right) ^{T}\mathbf{W}^{T}\left( \mathbf{R}-%
\mathbf{W}d^{true}+\mathbf{R}-\mathbf{W}\hat{d}\left( t_{2}\lambda \right)
\right) }{n} \\
&\geq &t_{2}^{2}-\left( \theta ^{ora}\right) ^{2}-\left\Vert \hat{d}\left(
t_{2}\lambda \right) -d^{true}\right\Vert _{1}\left( \nu +t_{2}\lambda
\right)  \\
&\geq &\left( t_{2}+\theta ^{ora}\right) \theta ^{ora}\tau -\frac{2\xi }{\xi
+1}t_{2}\lambda \left\Vert \hat{d}\left( t_{2}\lambda \right)
-d^{true}\right\Vert _{1} \\
&\geq &2\theta ^{ora}\left( \tau \theta ^{ora}-\frac{3\xi }{2\left( \xi
+1\right) }\lambda \left\Vert \hat{d}\left( t_{2}\lambda \right)
-d^{true}\right\Vert _{1}\right) >0\text{,}
\end{eqnarray*}%
where the second last inequality is due to the fact $\tau \leq 1/2$ and the
last inequality follows from the definition of $\tau $ and the $l_{1}$ error
bound in Equation (\ref{standard Lasso Result: L1 norm bound}) of
Proposition \ref{Standard Lasso Result}. Still, our choice of $\tau $ is
well-defined and is larger than $\frac{\lambda }{\theta ^{ora}}\frac{4\xi }{%
2\left( \xi +1\right) }\left\Vert d^{true}-\hat{d}\left( t_{2}\lambda
\right) \right\Vert _{1}$.

\section*{Acknowledgements}

We thank Yale University Biomedical High Performance Computing Center for
computing resources, and NIH grant RR19895 and RR029676-01, which funded the
instrumentation. The research of Zhao Ren and Harrison Zhou was supported in
part by NSF Career Award DMS-0645676, NSF FRG Grant DMS-0854975 and NSF
Grant DMS-1209191.

\bibliographystyle{ECA_jasa}
\bibliography{ConditionalGGM}

\end{document}